\documentclass[12pt]{article}

\pdfoutput=1

\usepackage{latexsym}
\usepackage{epsfig,amssymb,euscript}
\usepackage{amsmath}
\usepackage{mathrsfs}
\usepackage{ccaption}    
\usepackage[usenames,dvipsnames]{color}
\usepackage{enumerate}
\usepackage{ dsfont }

\usepackage[
      colorlinks=true,
      linkcolor=blue,
      urlcolor=blue,
      filecolor=black,
      citecolor=red,
      pdfstartview=FitV,
      pdftitle={},
        pdfauthor={Simon A. Gentle, Cynthia Keeler},
        pdfsubject={},
        pdfkeywords={},
        pdfpagemode={},
        bookmarksopen=true
      ]{hyperref}

  \captionnamefont{\bfseries}

\usepackage{subfigure}

\def\be{\begin{equation}}
\def\ee{\end{equation}}
\def\bea{\begin{eqnarray}}
\def\eea{\end{eqnarray}}

\def\domd{\lozenge_{\cal A}}  
\def\cwedge{\blacklozenge_{{\cal A}}} 

\def\bcwedgef{\partial_+(\blacklozenge_{{\cal A}})}
\def\bcwedgep{\partial_-(\blacklozenge_{{\cal A}})}

\numberwithin{equation}{section}

\newcommand{\m}{m}

\title{\bf On the reconstruction of Lifshitz spacetimes}

\author{\large Simon A.~Gentle$^a$ and Cynthia Keeler$^b$\\ \\
        \small \it $^a$\it Department of Physics and Astronomy \\
        \small \it University of California, Los Angeles, CA 90095, USA \\ \\
        \small $^b$\it Niels Bohr International Academy, Niels Bohr Institute \\
        \small \it University of Copenhagen, Blegdamsvej 17, DK 2100, Copenhagen, Denmark
 \\ \\
        \normalsize\href{mailto:sgentle@physics.ucla.edu}{\texttt{sgentle@physics.ucla.edu}}\texttt{, }\href{mailto:keeler@nbi.ku.dk}{\texttt{keeler@nbi.ku.dk}}}
\date{}

\begin{document}

\setlength{\baselineskip}{18pt}

\maketitle

\thispagestyle{empty}                        

\begin{abstract}
\setlength{\baselineskip}{18pt}
We consider the reconstruction of a Lifshitz spacetime from three perspectives: differential entropy (or `hole-ography'), causal wedges and entanglement wedges. We find that not all time-varying bulk curves in vacuum Lifshitz can be reconstructed via the differential entropy approach, adding a caveat to the general analysis of \cite{Headrick:2014eia}.  We show that the causal wedge for Lifshitz spacetimes degenerates, while the entanglement wedge requires the additional consideration of a set of boundary-emanating light-sheets.  
The need to include bulk surfaces  with no clear field theory interpretation in the differential entropy construction and the change in the entanglement wedge formation both serve as warnings against a naive application of holographic entanglement entropy proposals  in Lifshitz spacetimes.

\end{abstract}

\pagebreak

\tableofcontents


\section{Introduction}%
\label{sec:intro}

Holographic dualities purport to map field theories  to gravitational systems in higher dimensions.  The first examples of holographic duality mapped conformal (relativistic) field theories in $d$ spacetime dimensions to $(d+1)$-dimensional Anti-de Sitter space.  Interest in using the power of holography, as a strong-weak duality, for applications in condensed matter physics has driven  extensions of holographic dualities to non-relativistic field theories. 

If these maps are actually dualities, then they should run in both directions: that is, we should be able to reconstruct bulk information from boundary data, in addition to computing boundary quantities from bulk physics.  In the AdS context, boundary entanglement entropies have been successfully used to reconstruct information about the bulk spacetime metric. The work of this paper will be to extend  AdS  reconstruction schemes to Lifshitz spacetimes. 

Initial approaches to spacetime reconstruction concentrated on rebuilding field profiles on top of a fixed background; for example,  via the smearing function \cite{Balasubramanian:1999ri}.  This smearing function, which takes normalizable boundary information and outputs the bulk field profile, requires a momentum cut-off, limiting its locality, in both black hole spacetimes \cite{Freivogel:2013zta} and non-relativistic spacetimes including Lifshitz spacetimes \cite{Keeler:2013msa,Keeler:2014lia}.

To fully rebuild the spacetime, however, we must also reconstruct the bulk metric itself. Success at reconstructing metric information has increased since the Ryu-Takayanagi (RT) proposal \cite{Ryu:2006ef,Ryu:2006bv} and its subsequent covariant generalization by Hubeny-Rangamani-Takayanagi (HRT)  \cite{Hubeny:2007xt}  were developed.  These proposals relate the entanglement entropy of a spacelike co-dimension one  boundary region to the area of the bulk extremal surface homologous to this region. This relationship was  used by \cite{Lashkari:2013koa, Faulkner:2013ica}  to perturbatively reconstruct Einstein's equation linearized about  AdS spacetime.  Direct approaches to reconstruct the metric from boundary data include \cite{Hammersley:2006cp, Hammersley:2007ab, Bilson:2008ab, Bilson:2010ff}.  Additionally, the `hole-ography' \cite{Balasubramanian:2013lsa} and differential entropy \cite{Myers:2014jia, Headrick:2014eia} approaches reconstruct the length of a bulk curve from a family of boundary entanglement entropies.  

As the entanglement entropy is defined in terms of the density matrix for a boundary subregion, it is natural to ask what bulk region is reconstructible from this density matrix \cite{Czech:2012bh, Hubeny:2012wa, Headrick:2014cta}. Two regions to consider are the `entanglement wedge' \cite{Headrick:2014cta} and the `causal wedge' \cite{Hubeny:2012wa}, the latter of which additionally defines the causal holographic information surface.  The entanglement wedge is the causal development of the bulk slice bounded by the  extremal surface used to compute entanglement entropy and the boundary subregion itself.   The causal wedge, on the other hand, depends purely on  causal relations: it is  the intersection of the bulk past and bulk future of the boundary domain of dependence of the boundary subregion.

The particular class of non-relativistic field theories whose dual we will study are those obeying Lifshitz symmetry.  Under this symmetry, time and space scale differently: $ x \rightarrow \Lambda x$ while $t \rightarrow \Lambda^z t $, where $z$ is referred to as the dynamical exponent.  The dual spacetime, termed Lifshitz spacetime, was proposed in \cite{Kachru:2008yh, Taylor:2008tg} and has been studied intensively since (see \cite{Taylor:2015glc} for a recent review). In this paper we consider the reconstruction of a Lifshitz spacetime from three perspectives: differential entropy, causal wedges and entanglement wedges. 

Since entanglement-based spacetime reconstruction schemes often rely upon the Ryu-Takayanagi proposal and its covariant generalization, let us pause to consider its application to Lifshitz spacetimes. Most authors have assumed these proposals apply  to spacetimes with non-relativistic duals; for example, see \cite{Azeyanagi:2009pr, Solodukhin:2009sk, Keranen:2011xs, Kim:2012nb, Alishahiha:2014cwa, Fonda:2014ula, Fischetti:2014zja, Hosseini:2015gua} (and also \cite{Singh:2013iba, Singh:2013pfa}).  For  vacuum Lifshitz spacetime, the induced metric on a constant time slice   cannot be distinguished from a constant (Poincar\'{e}) time slice in AdS; this means that any result confined to a constant time slice in vacuum Lifshitz will be as if the dynamical exponent $z$ has been set to one. See \cite{Fradkin:2006mb, Solodukhin:2009sk, Hsu:2010ag, Oshikawa:2010kv, InglisMelko} for a sample of entanglement entropy calculations in Lifshitz field theories.

The simplest way to probe the dynamical exponent is to consider the entanglement entropy of a boundary region not on a constant time slice.  However, given both the lack of conformal boundary in a Lifshitz spacetime and the even more important lack of a physical interpretation for a  boundary region of non-constant time in a non-relativistic field theory, this approach cannot provide field theoretic justification for the covariant holographic entropy proposal in Lifshitz spacetimes.  Even for boundary regions on a constant time slice, where the dependence on $z$ is introduced via a non-zero temperature black brane setup or some other asymptotically Lifshitz solution, comparison with field theory results is difficult.  Lifshitz field theories have less symmetry than conformal field theories, so many of the field theory entanglement entropy calculation techniques do not work in this context.   Of course this difficulty highlights the importance of having a holographic entanglement calculation method, while also making such a method hard to verify.

For the body of this paper, our approach will also be to treat the RT and HRT entanglement entropy proposals as applicable to Lifshitz spacetimes.  However, as we discover in section \ref{sec:entanglementwedge} and discuss further in the conclusions, the peculiar nature of light-sheets in Lifshitz spacetime actually alters one of the justifications for the covariant entanglement entropy proposal.

There are two major differences between Lifshitz spacetimes and AdS that will underlie our bulk reconstruction difficulties.  First, only purely radially-directed light rays in Lifshitz reach the boundary.  Any light ray with nonzero momentum in a spatial direction will instead reach a turning point and then return to the bulk. This feature will complicate both curve reconstruction via differential entropy and the construction of the entanglement wedge.

The second major difference is the degeneracy of the Lifshitz boundary.  Unlike in AdS, Lifshitz does not have a conformal boundary; the metric on the boundary is degenerate. This degeneracy means the boundary domain of dependence collapses, due to the non-relativistic nature of the boundary.  Since it is non-relativistic, the effective speed of light is infinite, so any point in the spacetime can be affected by any point at an earlier time, regardless of its  location in space. In fact, since the boundary is non-relativistic, we expect a metric-complex instead of a boundary metric.  This vielbein approach to defining the boundary has been successful for holographic renormalization in non-relativistic duals \cite{Horava:2009vy, Ross:2009ar, Ross:2011gu, Chemissany:2014xsa, Andrade:2014iia, Andrade:2014kba,Hartong:2014oma,Hartong:2014pma,Christensen:2013lma,Christensen:2013rfa}.   Of course, we can avoid these issues if we work at a cut-off. However, we expect that removing the cut-off will result in degeneration of the causal wedge; this is in fact the behavior we see in section \ref{sec:causalwedges}.

This paper is organized as follows.  In the following section we collect conventions and simple observations about Lifshitz spacetimes that will be useful throughout our discussion. In section \ref{reconstruction} we obtain partial success in the reconstruction of bulk curves via the differential entropy approach. We discuss the details of the relationship between curves that are not reconstructible and light rays that turn around in Lifshitz in appendix~\ref{appendix:screens}. Section \ref{sec:causalwedges} studies the degeneration of the causal wedge, while in section \ref{sec:entanglementwedge} we construct the entanglement wedge.  Section \ref{sec:Discussion} contains discussion of our results and speculations about how to  better reconstruct a  spacetime dual to a Lifshitz field theory.

\section{Geodesics in Lifshitz spacetime}%
\label{sec:conventions}

We work with the Lifshitz metric
\begin{equation}\label{eq:Lifshitzmetric}
ds^2 = L^2\left(-\frac{dt^2}{u^{2 z}} +  \frac{dx^2}{u^2}+ \frac{du^2}{u^2} \right)
\end{equation}
where the boundary is at $u=0$.  Although we will work almost entirely in three dimensions, the majority of our results will hold when $dx^2$ is replaced by $d\vec{x}^2$.

This metric coincides with AdS$_3$ in Poincar\'{e} coordinates when $z=1$.  We will only consider $z\geq 1$ as this restriction satisfies the null energy condition \cite{Hoyos:2010at}.  As in AdS, $L$ sets the scale for the radius of curvature of the geometry, though we will  set this to unity from now on for simplicity.   In section~\ref{reconstruction} we will demand that the $x$ direction is periodic with period $\xi$. 

The radial motion of  geodesics in the geometry \eqref{eq:Lifshitzmetric} (with $L=1$) is governed by an effective potential as follows:
\begin{equation}\label{eq:gengeodesics}
\dot u^2 = - V_{\textrm{eff}}^{(\kappa)}(u) \equiv   \kappa\, u^2 +  E^2\, u^{2(z+1)}  - P^2\, u^4.
\end{equation}
A dot denotes derivative with respect to an affine parameter and $\kappa=1,0$ or $-1$ for spacelike, null or timelike geodesics, respectively. The conserved quantities $E$ and $P$ are associated respectively with the Killing vectors $\partial_t$ and $\partial_x$ and satisfy
\begin{align}
\label{eq:Edef}
\dot{t} 
&= E u^{2z},
\\
\label{eq:Pdef}
\dot{x}
&= P u^2.
\end{align}
In section~\ref{reconstruction} we will be concerned with spacelike geodesics and the reconstruction of bulk curves.

\begin{figure}[hb!]
\begin{center}
\includegraphics[width=0.45\textwidth]{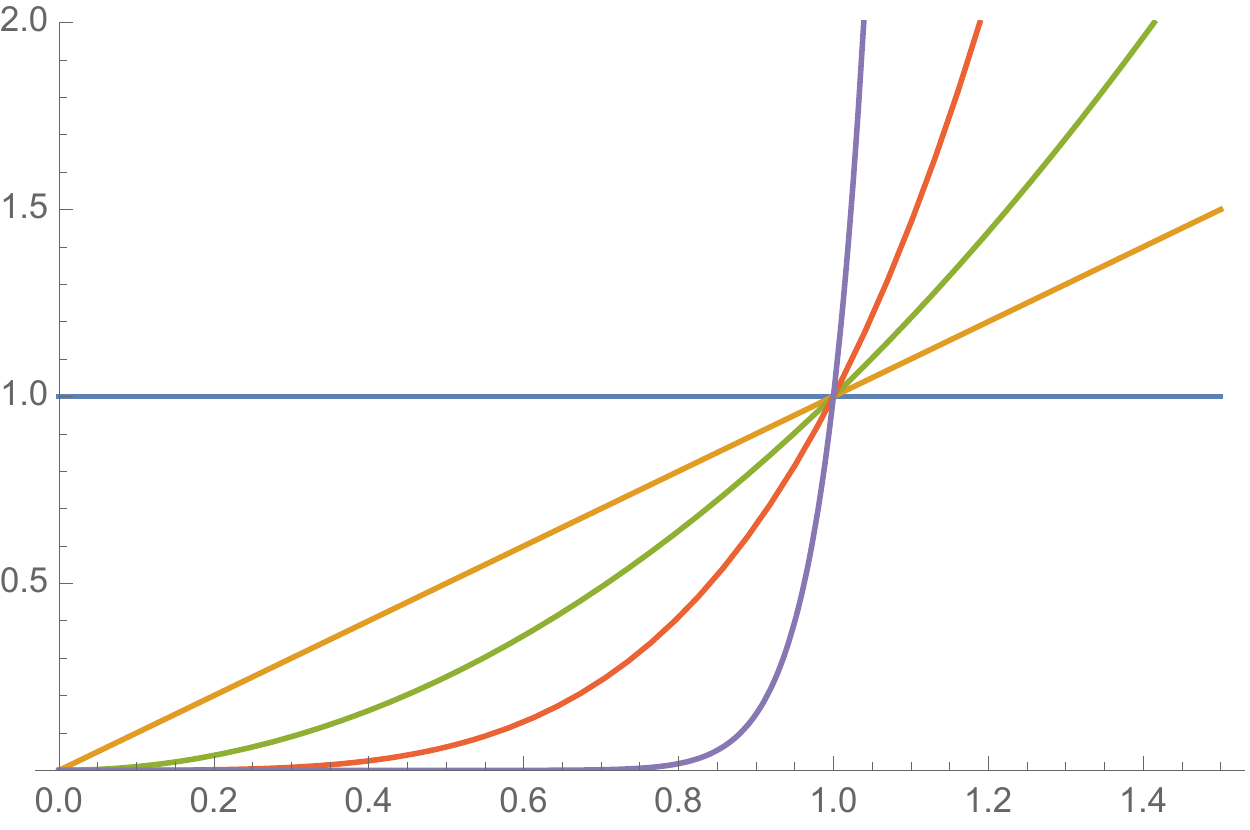}
\setlength{\unitlength}{0.1\columnwidth}
\begin{picture}(0.3,0.4)(0,0)
\put(-5.,2.8){\makebox(0,0){$\ell^2$}}
\put(0.2,0.2){\makebox(0,0){$u$}}
\put(-3.6,1.2){\makebox(0,0){increasing $z$}}
\put(-2.85,1.05){\vector(1,-1){0.55}}
\end{picture}
\end{center}
\caption{Boundaries  $\ell^2 =  u^{2(z-1)}$ below which the radial effective potential for null geodesics $V_{\textrm{eff}}^{(0)}(u)< 0$ for different values of $z$: 1 (blue), 3/2 (yellow), 2 (green), 3 (red)  and 10 (purple).}\label{fig:potentialLifshitz}
\end{figure}
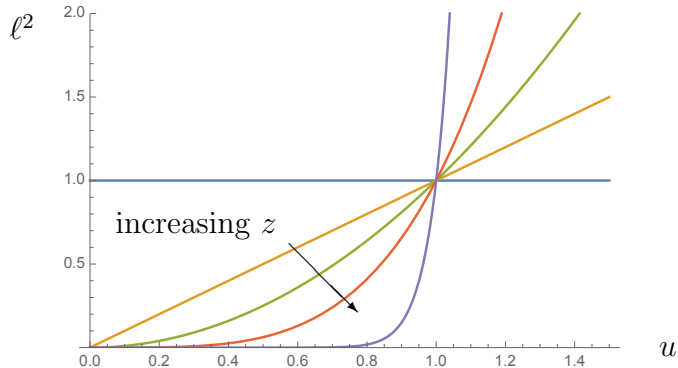

In section~\ref{sec:causalwedges} we will focus on null geodesics. Beginning with the general geodesic equation \eqref{eq:gengeodesics}, we set $\kappa=0$, rescale the affine parameter by $E$ and introduce the  (rescaled) transverse momentum $\ell\equiv P/E$:
\begin{equation}\label{eq:nullgeodesics}
\dot t = u^{2 z}, \quad \dot u^2=-V_{\textrm{eff}}^{(0)}(u) \equiv u^4 \left( u^{2 (z-1)} - \ell^2 \right),\quad \dot x = \ell u^2.
\end{equation}
As  pointed out in \cite{Keeler:2013msa},  light rays  with transverse momentum do not reach the boundary in Lifshitz spacetime.  This phenomenon is manifested by a bump in the potential $V_{\textrm{eff}}^{(0)}(u)$ near the boundary for any non-zero $\ell$.  Classical geodesic motion must obey $\dot u^2\geq 0$ and therefore can only take place in regions for which $V_{\textrm{eff}}^{(0)}(u)\leq 0$, i.e.\ for $u\geq  |\ell|^{1/(z-1)}$. The boundaries of these regions for different values of $z$ are plotted in figure~\ref{fig:potentialLifshitz}. By contrast, light rays reach the boundary of an  asymptotically AdS spacetime if $\ell^2 \leq 1$ (using the same conventions).

For large $z$, spacetime effectively cuts off  below $u= 1$ as far as null geodesics with $\ell \ne 0$ are concerned.   This is apparent even at $z=10$ for the purple curve in figure~\ref{fig:potentialLifshitz}. Since  Lifshitz spacetime becomes AdS$_2 \times \mathbb{R}^{D-2}$ as $z\rightarrow \infty$, this behavior is expected.  Light rays with non-zero momentum in the flat directions are headed towards the flat space's asymptotic infinity instead of the conformal boundary of AdS$_2$.

\section{Reconstructing bulk surfaces}%
\label{reconstruction}

Consider a bulk surface not anchored on the boundary of a holographic spacetime. Assuming this bulk surface is not a horizon, what is its boundary interpretation?  Can it be reconstructed from boundary observables?  In this section we address these questions for Lifshitz spacetime.

Building on the work of \cite{Balasubramanian:2013lsa, Myers:2014jia},  a general proof relating the surface area of a generic class of bulk surfaces to the `differential entropy' of a particular family of boundary intervals was given in \cite{Headrick:2014eia}.  These boundary intervals are chosen so that the extremal surfaces that calculate their holographic entanglement entropy, via the HRT conjecture, are tangent to the bulk surface whose area is being computed. In section~\ref{sec:differentialentropy} we review this construction and consider an explicit example in Lifshitz spacetime.

However, despite the success of section~\ref{sec:differentialentropy}, not all curves in Lifshitz spacetime can be constructed via differential entropy, as we exhibit explicitly for a family of bulk curves in section~\ref{sec:badcurves}.  For these curves, there are some regions where no tangent extremal surface can reach the boundary.  The situation is reminiscent to that found in \cite{Engelhardt:2015dta}, which showed that spacetimes with holographic screens, such as black hole spacetimes, must contain bulk curves beyond said screens that are not reconstructible via the `hole-ography' or differential entropy approach.

\subsection{Holographic holes and differential entropy}
\label{sec:differentialentropy}

For simplicity we begin our discussion in Poincar\'{e} AdS$_3$, given in \eqref{eq:Lifshitzmetric} with $z=1$.  In this section we take the $x$ direction to be periodic with period $\xi$. 

Consider a closed bulk curve $\gamma_B(\lambda)$ that is not  anchored on the boundary. This must respect the periodicity in $x$ and we insist that its tangent vector be everywhere spacelike, but  otherwise it can be arbitrary.  In particular, it need not be extremal.  Its area, or `gravitational entropy', is given by the Bekenstein-Hawking formula:
\begin{equation}\label{eq:gravitationalentropy}
S_G \equiv \frac{1}{4 G_N} \int_0^1 d\lambda\, \left| \gamma'_B \right| ,
\end{equation}
where a prime denotes differentiation with respect to $\lambda$.

The construction of \cite{Headrick:2014eia} is built around a family of boundary intervals with endpoints described by the two curves $\gamma_\pm^\m(\lambda)$ (where $m\in\{t,x\}$).  These curves depend periodically on the parameter $\lambda$ such that the intervals cover a full period of $x$ as $\lambda$ varies in $[0,1]$.  Although each interval must lie on a boundary Cauchy slice, they need not all lie on the same one.  From the curves $\gamma_\pm^\m(\lambda)$, we construct a continuous family of bulk spacelike geodesics  $\Gamma(s;\lambda)$, where $\lambda$ labels the geodesic and $s$ is a  parameter (not necessarily affine) along it.  The  endpoints of geodesic $\lambda$ are anchored  on the boundary at $\gamma_\pm^\m(\lambda)$.  As in the HRT prescription, if there is more than one geodesic anchored at a particular pair of boundary endpoints, we take the  shortest such geodesic. 

The  differential entropy of this family of boundary intervals is defined by
\begin{equation}\label{eq:differentialentropy}
S_D \equiv \int_0^1 d\lambda\, \left.\frac{\partial S[\gamma_-(\lambda),\gamma_+(\lambda')]}{\partial \lambda'} \right|_{\lambda'=\lambda}.
\end{equation}
In this formula, $S[\gamma_-(\lambda),\gamma_+(\lambda)]$ is the entanglement entropy associated with the particular boundary interval at $\lambda$.  This is given holographically by \cite{Hubeny:2007xt} as the length (divided by $4 G_N$) of the bulk spacelike geodesic $\Gamma(s;\lambda)$.  

We have suggestively used $\lambda$ to parametrize both the bulk curve $\gamma_B$ and the family of boundary intervals described by the boundary curves $\gamma_\pm^\m$, because, as proven in \cite{Headrick:2014eia}, the gravitational entropy and the boundary differential entropy are the same.  That is, $S_G=S_D$ if we demand that the geodesics $\Gamma(s;\lambda)$ are tangent to the bulk curve $\gamma_B(\lambda)$ at some $s=s_B(\lambda)$.  This requires
\begin{align}
\Gamma(s_B(\lambda);\lambda) &= \gamma_B(\lambda) \label{eq:TVA1}\\
\dot{\Gamma}(s_B(\lambda);\lambda) &= \alpha(\lambda)\, \gamma_B'(\lambda) \label{eq:TVA2},
\end{align}
where a dot denotes differentiation with respect to $s$. 

Before we proceed, note that \cite{Headrick:2014eia} generalized this result; for example, by relaxing the condition \eqref{eq:TVA2} to allow the tangent vectors to be oppositely-oriented or not even collinear.  In our simple example we will take $\alpha(\lambda)>0$. 

The work in  \cite{Headrick:2014eia} also extends to a variety of other backgrounds that possess planar symmetry,  including Lifshitz spacetime.%
\footnote{One can also extend the result to include spacetimes in higher dimensions with a generalised notion of `planar symmetry'.  This effectively means that the additional dimensions can be `factored out' and intervals are replaced with strips. We consider three bulk dimensions for simplicity.}
The relationship between the lengths of the boundary-anchored extremal curves $\gamma^m_\pm(\lambda)$ and the length of the bulk curve $\gamma_B$ is clearly preserved in Lifshitz spacetime.  However, in order to actually relate the differential entropy to the gravitational entropy, we require that the entanglement entropy in the boundary theory is still computed holographically by \cite{Hubeny:2007xt}.   It is not clear that this is the case for Lifshitz spacetime; for instance, the interpretation for the length of a bulk spacelike geodesic with endpoints at different times is unknown.  The purpose of this section is to demonstrate the extent to which the geometric construction still goes through, regardless of the entropic interpretations, whilst  also highlighting the differences between a Lifshitz and an asymptotically AdS spacetime.

Now we turn to Lifshitz spacetime \eqref{eq:Lifshitzmetric} with $z>1$.  Consider a bulk curve that sits at a fixed radius $u=u_{\star}$ but varies in time and respects the periodicity in $x$.  We parametrize this curve by
\begin{gather}
\gamma_B(\lambda) = \{ T(\lambda), \xi\, \lambda, u_{\star} \} \quad \textrm{with} \quad \lambda\in[0,1] \label{eq:gammaB},
 \\
 \gamma_B(0) = \gamma_B(1) \label{eq:periodicBCgammaB}.
\end{gather}
Since we want the bulk curve to be spacelike, its tangent vector must be everywhere spacelike:
\begin{equation}\label{eq:spacelikecondition}
T'(\lambda)^2 < \xi^2\, u_{\star}^{2(z-1)}.
\end{equation}
We can immediately write down length of this curve:
\begin{equation}\label{eq:gravitationalentropy2}
S_G = \frac{1}{4 G_N} \int_0^1 d\lambda\, \sqrt{- \frac{T'(\lambda)^{2}}{u_{\star}^{2z}} + \frac{\xi^2}{u_{\star}^2} }.
\end{equation}

We need a continuous family of bulk spacelike geodesics  that begin and end at the boundary $u=0$ in order to construct the differential entropy.  
The equations we must solve are \eqref{eq:gengeodesics}, \eqref{eq:Edef} and \eqref{eq:Pdef}.  It is convenient to parametrize the spacelike geodesics via the radius $u$.  Consequently, we find pairs of  functions $t_\pm(u)$ and $x_\pm(u)$ that describe the two halves of a geodesic, one on each side of its radial turning point. These two halves must match smoothly at the bulk curve $\gamma_B(\lambda)$ so that the resulting geodesics are tangent to the curve, i.e.\ so  \eqref{eq:TVA1} and \eqref{eq:TVA2} are satisfied.%
\footnote{Note that in the case of $E=0$ the geodesic is restricted to lie on a constant-time slice.  In this case the problem is identical to that studied in section 2.3 of \cite{Headrick:2014eia} (in fact, for any $z$).  This geodesic will also be useful later in sections~\ref{sec:causalwedges} and \ref{sec:entanglementwedge}.}

The matching point between the two portions of the geodesic is where it touches the bulk curve, at $u=u_{\star}$. From  \eqref{eq:TVA1} we have 
\begin{align}
t_-(u_{\star}) &= t_+(u_{\star}) = T(\lambda) \label{eq:tmatching}, 
\\
 x_-(u_{\star}) &= x_+(u_{\star}) = \xi\, \lambda \label{eq:xmatching}.
\end{align}
This is also the turning point of the geodesic, where
\begin{equation}\label{eq:generalturningpoint}
\left. \dot{u}\right|_{u= u_{\star}} =0 \quad \Rightarrow \quad V_{\textrm{eff}}^{(\kappa=1)}(u_{\star})=0.
\end{equation}
The ratio $\dot{t}/\dot{x}$ is fixed in terms of the geodesic's conserved quantities $E$ and $P$  via  \eqref{eq:Edef} and \eqref{eq:Pdef}:
\begin{equation}\label{eq:tdotoverxdot}
\left.\frac{\dot{t} }{\dot{x} }\right|_{u= u_{\star}}  = \frac{E\, u_{\star}^{2(z-1)}}{P} = \frac{T'(\lambda)}{\xi},
\end{equation}
where we have applied the second tangency condition \eqref{eq:TVA2} for the right hand equality.  We also note that the spacelike condition (\ref{eq:spacelikecondition}) on the bulk curve can now be rewritten as
\begin{equation}\label{eq:spacelikePE}
u_\star ^{2(z-1)} < \frac{P^2}{E^2} .
\end{equation}
That is, a bulk curve at constant radius $u_\star$ with a tangent geodesic having conserved quantities $E,P$ is spacelike as long as the bulk curve is located at a radius smaller then the maximum in (\ref{eq:spacelikePE}).  Since the bulk curve's tangent geodesics could have different conserved quantities at different tangent points, the maximum possible radius is set by the smallest $P/E$ attained along the entire bulk curve.

Moving forward, we  evaluate $E,P$  using the chain rule and taking the limit $u\to u_{\star}$:
\begin{align}
E &= \beta\, \frac{T'(\lambda)}{u_{\star}^z \sqrt{\xi^2\, u_{\star}^{2(z-1)}-T'(\lambda)^2} }  \label{eq:Eoflambda}, \\
P &= \beta\, \frac{\xi\, u_{\star}^{z-2}}{ \sqrt{\xi^2\, u_{\star}^{2(z-1)}-T'(\lambda)^2} } \label{eq:poflambda},
\end{align}
where $\beta=\pm 1$.  Comparing this result with \eqref{eq:TVA2} we  identify
\begin{equation}
\alpha(\lambda)  \equiv \beta\, \frac{u_{\star}^{z}}{ \sqrt{\xi^2\, u_{\star}^{2(z-1)}-T'(\lambda)^2} }.
\end{equation}
We will choose $\beta=1$ so that the orientation of the two tangent vectors $\dot\Gamma$ and $\gamma_B'$ agree at the point $u=u_\star$.%
\footnote{The opposite choice is perfectly valid. In that case the differential entropy computes the `signed length' of the curve: \eqref{eq:gravitationalentropy} supplemented with $\textrm{sgn}\, \alpha$ \cite{Headrick:2014eia}.}

From here on we restrict to $z=2$ for which we can  solve the equations  \eqref{eq:gengeodesics}, \eqref{eq:Edef} and \eqref{eq:Pdef} analytically.   The turning point in this case satisfies 
\begin{equation}\label{eq:ustar}
1-P^2 u_{\star}^2 + E^2 u_{\star}^{4} = 0 \quad \Rightarrow \quad u_{\star} = \frac{\sqrt{P^2 - \sqrt{P^4-4 E^2} }}{\sqrt{2}E}.
\end{equation}
Here we have chosen the smallest  positive root because we require boundary-anchored geodesics, as we will explicitly demonstrate below. This root is real as long as the bulk curve's tangent geodesics have $P^4 - 4 E^2 \geq 0$. 
However, choosing the smallest positive root does have a consequence: the maximum turning radius we can produce from this smallest root is
\begin{equation}\label{eq:umax}
u_{\star,\textrm{max}}^2 = \frac{P^2}{2 E^2}.
\end{equation}
The astute reader will notice that this maximum radius is smaller than the maximum in \eqref{eq:spacelikePE}.  Regardless, we now impose the smaller maximum, and show that reconstruction works for bulk curves whose tangent geodesics and radius satisfy $u_\star \leq  u_{\star,\textrm{max}}$.  We will return to the case of larger radius in section~\ref{sec:badcurves} below.

The solutions that satisfy the matching conditions \eqref{eq:tmatching} and \eqref{eq:xmatching} are given by
\begin{align}
t_{\pm}(u;\lambda) &=  T(\lambda)  \pm \Delta t \pm \left[ \frac{1}{2E} ( 1- \sqrt{1-P^2 u^2 + E^2 u^{4} } ) \right. \nonumber
\\
&\phantom{=\ }\left. + \frac{P^2}{4 E^2} \log \left( \frac{ P^2-2E}{ P^2- 2E(E u^2 + \sqrt{1-P^2 u^2 + E^2 u^{4} })}   \right) \right] ,
\\
 x_{\pm}(u;\lambda) &=  \xi\, \lambda \pm \Delta x \pm \frac{P}{2 E} \log \left( \frac{ P^2-2E}{ P^2- 2E(E u^2 + \sqrt{1-P^2 u^2 + E^2 u^{4} })}  \right) .
\end{align}
where  we have defined 
\begin{align}
 \Delta t &\equiv -\frac{1}{2E}- \frac{P^2}{4E^2}\log \left( \frac{ P^2-2 E}{  \sqrt{P^4-4 E^2 }}   \right)   \label{eq:Deltat},
 \\
 \Delta x &\equiv -  \frac{P}{2E}\log \left( \frac{ P^2-2 E}{  \sqrt{P^4-4 E^2 }}    \right)  \label{eq:Deltax}.
\end{align}
As we can see from these explicit expressions,  these geodesics do indeed reach the boundary $u=0$.  In fact, from the endpoints at $u=0$  we can read off the family of boundary intervals necessary to reconstruct the bulk curve:
\begin{equation}
\gamma_\pm^{\m} = \{ T(\lambda) \pm \Delta t , \xi\, \lambda \pm \Delta x\}  \label{eq:endpoint}.
\end{equation}

As in an asymptotically AdS spacetime, the length of such a boundary-anchored geodesic diverges in Lifshitz spacetime.  We  introduce a simple radial cut-off at $u=\varepsilon$ to regulate this divergence:
\begin{align}
{\cal L} &\equiv 2 \int_{\varepsilon}^{u_{\star}} \frac{du}{u \sqrt{1-P^2 u^2 + E^2 u^{4} }} \label{eq:lengthresult}
 \\
  &=\left. \log\left( \frac{u^2}{2-P^2 u^2 + 2 \sqrt{1-P^2 u^2 + E^2 u^4} } \right) \right|_{\varepsilon}^{u_{\star}} 
\nonumber  \\
  &= \log\left( \frac{4}{\sqrt{P^4 - 4 E^2}\, \varepsilon^2 } \right) + O(\varepsilon^2) \label{eq:lengthresultEp}.
 \end{align}
 where we used \eqref{eq:ustar} to obtain the final line. 
 
 It is not clear how to interpret the length of this geodesic in the dual field theory. As mentioned earlier, when $E=0$ we simply recover the AdS result, so it is tempting to identify this as the entanglement entropy \emph{\`{a} la} Ryu-Takayanagi.  This independence from $z$ is a bit surprising in itself. There is some supporting evidence from field theory calculations  featuring Lifshitz symmetry: both \cite{Fradkin:2006mb} and \cite{Solodukhin:2009sk}   recover an area law, though the former observes an additional sub-leading divergence.  However, these two different setups both have more symmetry than we do.  
 
 Even aside from this possible concern, we are left with a further question:  when the interval  \eqref{eq:endpoint} does not lie on a constant-time slice, what are we computing? We do not have a physical understanding of this boundary quantity.  For both of these reasons, it is still unclear what is the meaningful entanglement entropy calculation in a Lifshitz field theory.  For now we simply assume that the length of this geodesic computes the entanglement entropy according to $S\equiv {\cal L}/(4 G_N)$.   We will see that the regulatory cut-off drops out in this construction.

We now have all the ingredients we need to evaluate the differential entropy,  defined in \eqref{eq:differentialentropy}.  It is useful to rewrite this as 
\begin{equation}\label{eq:differentialentropy2}
\int_0^1 d\lambda \left.\frac{\partial S[\gamma_-(\lambda),\gamma_+(\lambda')]}{\partial \lambda'} \right|_{\lambda'=\lambda} = \int_0^1 d\lambda\, \frac{\partial S(\gamma_-,\gamma_+)}{\partial \gamma_+^{\m}}\, \frac{d \gamma_+^{\m}}{d\lambda} .
\end{equation}
We know $S(E,P)$, but would need to invert \eqref{eq:Deltat} and \eqref{eq:Deltax} to obtain $S(\gamma_-,\gamma_+)$.  However, since we only need its derivative in \eqref{eq:differentialentropy2}, we can use the chain rule and implicit differentiation.  From translation invariance in $t$ and $x$ we can express the length as $S(\Delta t,\Delta x)$. We evaluate the partial derivatives the hard way using \eqref{eq:lengthresultEp},  \eqref{eq:Deltat} and \eqref{eq:Deltax}:
\begin{equation}
\frac{1}{2}\,  \frac{\partial S}{\partial \Delta t}   = - \frac{E}{4 G_N} \quad \textrm{and} \quad \frac{1}{2}\,   \frac{\partial S}{\partial \Delta x}  = \frac{P}{4 G_N}.
\end{equation}
Our results  seem  surprisingly simple  at first.  However, as pointed out in the proof given by  \cite{Headrick:2014eia}, the simple explanation is that these are  the Hamilton-Jacobi equations from $S= {\cal L}/(4 G_N)$.  We  now use \eqref{eq:Eoflambda} and \eqref{eq:poflambda} to express our derivative as a function of $\lambda$:
\begin{align}
\frac{\partial S(\gamma_-,\gamma_+)}{\partial \gamma_+^{\m}}\, \frac{d \gamma_+^{\m}}{d\lambda} &= -\frac{E}{4 G_N}\, \frac{d \gamma_+^{t}}{d\lambda} + \frac{P}{4 G_N}\,  \frac{d \gamma_+^{x}}{d\lambda} \nonumber
 \\
 &= \frac{1}{4 G_N}\, \frac{1}{u_{\star}^2} \left[ \sqrt{\xi^2 u_{\star}^2-T'(\lambda)^2} +\frac{1}{2}\, \frac{d}{d\lambda} \log \left( \frac{\xi^2 u_{\star}^2-T'(\lambda)^2}{\xi^2 u_{\star}^2-2T'(\lambda)^2} \right) \right].
\end{align}
The total derivative term drops out of the integral in \eqref{eq:differentialentropy2} due to the periodic boundary conditions \eqref{eq:periodicBCgammaB} on $\gamma_B(\lambda)$.  Thus, we do indeed recover the length, or gravitational entropy \eqref{eq:gravitationalentropy2}, of the bulk curve.

In conclusion, the differential entropy construction worked just as in asymptotically AdS spacetime, despite the lack of field theory interpretation for the length of a geodesic with endpoints at different times.  Furthermore, the cut-off we introduced dropped out.

Whilst we considered the simplest non-trivial example of a time-dependent curve, the  general result of \cite{Headrick:2014eia} will still hold for some more general bulk curves.  However, in the following section we will see that there exist curves for which it does not.  In \cite{Headrick:2014eia}  it is also claimed that the bulk curve emerges from the intersection of neighboring entanglement wedges.  We will comment on this in the context of Lifshitz spacetime in section~\ref{sec:Discussion}.  

\subsection{A Lifshitz failure of the differential entropy reconstruction}
\label{sec:badcurves}

Despite the success of the differential entropy construction for curves at the constant radii given in (\ref{eq:ustar}), we will now show that some bulk curves cannot be built in this manner, because the extremal surfaces necessary to do so never reach the boundary.  Thus, they have no endpoints on the boundary and cannot trace out boundary curves from which to define the differential entropy.

The simplest such bulk curves are as  above:  they have constant radius but can vary in time as we move along the $x$ direction, as long as they stay spacelike.   We again want to find a family of spacelike geodesics tangent to the curve, necessarily at the radial turning points of the geodesics.  To be specific, equations (\ref{eq:gammaB}) through (\ref{eq:poflambda}) still apply.

However, as we highlighted above, in the case of $z=2$ the maximum radius bulk curve we can describe using the smallest positive root in the turning point condition is  (\ref{eq:umax}), while requiring the bulk curve to be spacelike only necessitates  (\ref{eq:spacelikePE}).
There is a mismatch here: what about bulk curves of constant radius between $P/\sqrt{2} E$ and $P/E$?\footnote{More properly we should translate back to the bulk curve quantities using (\ref{eq:Eoflambda}) and (\ref{eq:poflambda}), so instead we are worried about bulk curves at constant radius that have at least one tangent with $P/\sqrt{2} E$ smaller than their constant radius $u_\star$.}
Clearly there are tangent spacelike geodesics, because (\ref{eq:spacelikePE}) is satisfied.  So what goes wrong with (\ref{eq:umax})?  That maximum radius was found assuming we took the smallest positive root of the turning point equation for the tangent geodesic.  If we instead take the larger positive root, the maximum radius now becomes just $P/E$.  At any larger radius, a curve with tangents governed by $E,P$ would be timelike so not within our construction.

However, we chose the smallest positive root because those geodesics start and end on the spacetime boundary.  A geodesic whose turning point occurs at the larger positive root instead ends at a singularity: either the tidal singularity in vacuum Lifshitz, or a black hole singularity.  Importantly, spacelike geodesics that do not end on the boundary cannot participate in the differential entropy construction. 

To check that geodesics whose turning points occur at larger radii than in (\ref{eq:umax}) do not reach the boundary, we consider the concavity of the geodesic, given by $\ddot{u}$.  When $\ddot{u}<0$ at a turning point, the geodesics bends towards the boundary there.  When $\ddot{u}>0$ at a turning point, then instead the geodesics return to larger radii. To find $\ddot{u}$ at a generic turning point, we take the derivative of (\ref{eq:gengeodesics}) with respect to the affine parameter, finding
\begin{equation}
\ddot{u} = (z+1) E^2 u^{2z+1}+u-2 u^3 P^2,
\end{equation}
where we have restored general $z$.  Next, we use the turning point equation $\dot{u}=0$ to eliminate either $E$ or $P$ from the expression $\ddot u<0$.  We  find
\begin{align}
  u_\star^2 &< \frac{z}{(z-1) P^2}
\\
  u_\star^{2z} &< \frac{1}{(z-1) E^2}.
\end{align}
Of course these are actually the same maximum possible $u_{\star,\textrm{max}}$, just expressed in a different way.  Consequently, although earlier we only exhibited the mismatch for $z=2$, we now see the maximum radius for boundary-anchored spacelike geodesics for general $z$ obeys
\begin{equation}\label{eq:umaxgenz}
u_{\star,\textrm{max}}^{2(z-1)} = \frac{P^2}{z E^2}.
\end{equation}
Again, spacelike curves with $E,P$ exist up to $u_\star$ satisfying (\ref{eq:spacelikePE}), but for these larger radii their tangent geodesics never reach the boundary.

This failure of the differential entropy reconstruction is reminiscent of that shown in \cite{Engelhardt:2015dta}. The authors proved that spacetimes with certain types of  `holographic screens' contain bulk surfaces that cannot be reconstructed via this hole-olography approach (even though extremal surfaces can reach the bulk surfaces in question).  As we exhibit in appendix \ref{appendix:screens}, Lifshitz spacetime contains surfaces similar to holographic screens; these surfaces, caused because light rays with non-zero momentum turn around in Lifshitz spacetimes, are in fact indicative of the explicit failure we have just shown.

Our situation is actually slightly worse than that shown in the horizon-having spacetimes of \cite{Engelhardt:2015dta}.  In Lifshitz spacetime, by choosing a spacelike curve whose tangent somewhere has arbitrarily small $P/E$, we can draw a bulk spacelike curve arbitrarily close to the boundary whose tangent geodesics will never reach the boundary.  There are non-reconstructible constant radius bulk curves everywhere in the spacetime.

\section{Causal wedge degeneration}%
\label{sec:causalwedges}

In this section we consider a construction based on a proper subset of (a Cauchy slice of) the boundary: the causal wedge.

Consider first an asymptotically AdS spacetime and focus on a spacelike co-dimension one boundary region ${\cal A}$ with boundary domain of dependence $\domd$.  The bulk causal wedge $\cwedge$  is the intersection of the causal past and the causal future of $\domd$.  Its  boundary in the bulk  is generated by null geodesics that end on the boundary of $\domd$.  Given the knowledge of the reduced density matrix in ${\cal A}$, it was argued in \cite{Hubeny:2012wa} that one should be able to reconstruct at least the corresponding bulk causal wedge in the dual spacetime

In an asymptotically AdS spacetime, we only have to send null geodesics from the future- and past-most tips of $\domd$  in order to obtain a wedge that is a closed subset of the bulk.  As mentioned in the introduction, the boundary and causal structures of Lifshitz spacetime differ from those in AdS, so we will have to modify the above definitions appropriately.

\subsection{Causal wedges in AdS spacetimes}
\label{sec:AdSwedges}

\begin{figure}[h!]
\begin{center}
\hskip.75em
\includegraphics[width=0.45\textwidth]{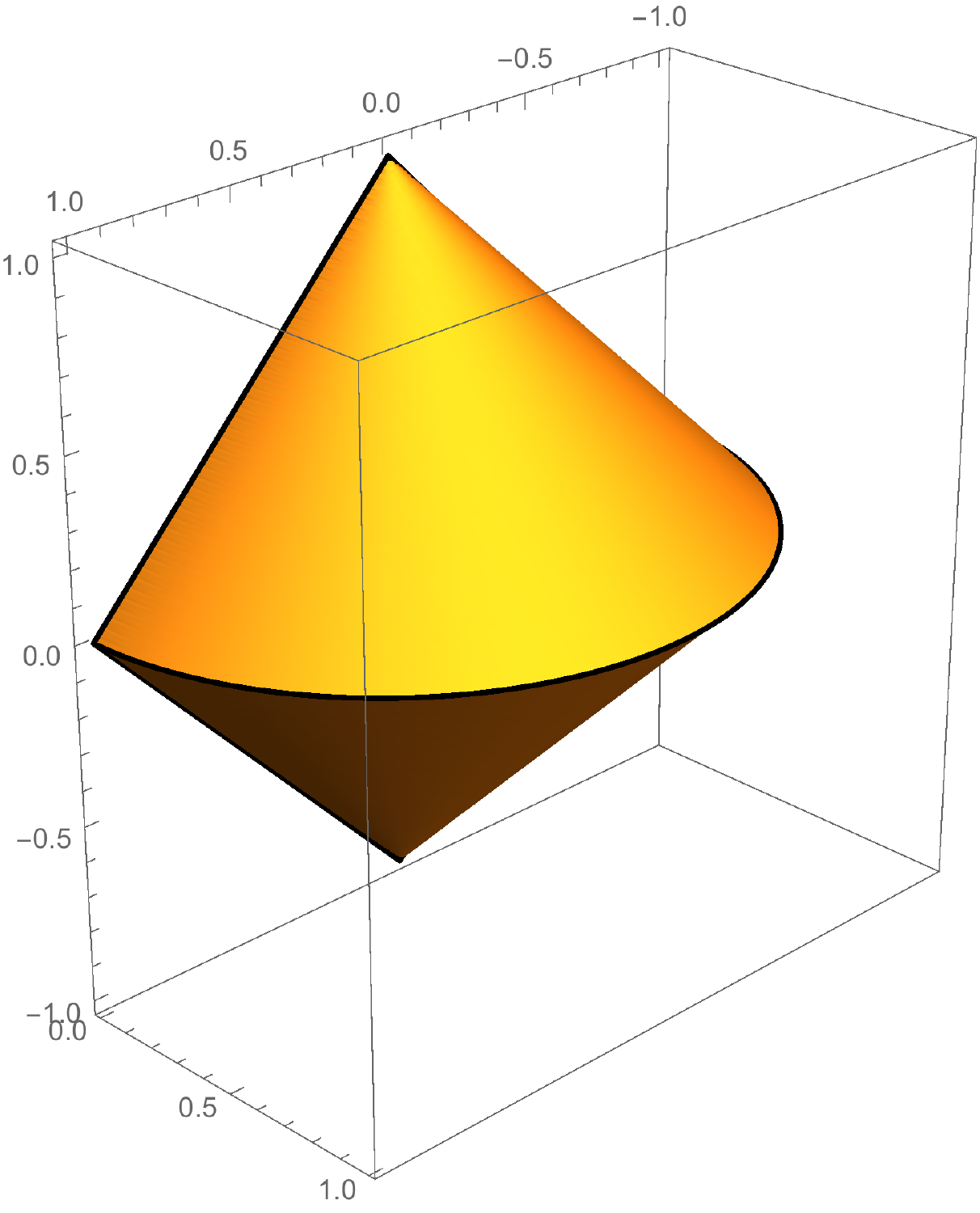}
\setlength{\unitlength}{0.1\columnwidth}
\begin{picture}(0.3,0.4)(0,0)
\put(-4.1,4.2){\makebox(0,0){$\domd$}}
\put(-3.85,3.95){\vector(1,-1){0.25}}
\put(-0.5,1.8){\makebox(0,0){$\Xi_{\cal A}$}}
\put(-0.8,2){\vector(-1,1){0.65}}
\end{picture}
\end{center}
\caption{The causal wedge $\cwedge$ in  Poincar\'{e} AdS$_3$. We fix the interval width $2a=2$ and also show the causal  information surface $\Xi_{\cal A}$ (given by \eqref{eq:futurewedgebdyAdS} with $t=0$).}\label{fig:causalwedgeAdS}
\end{figure}
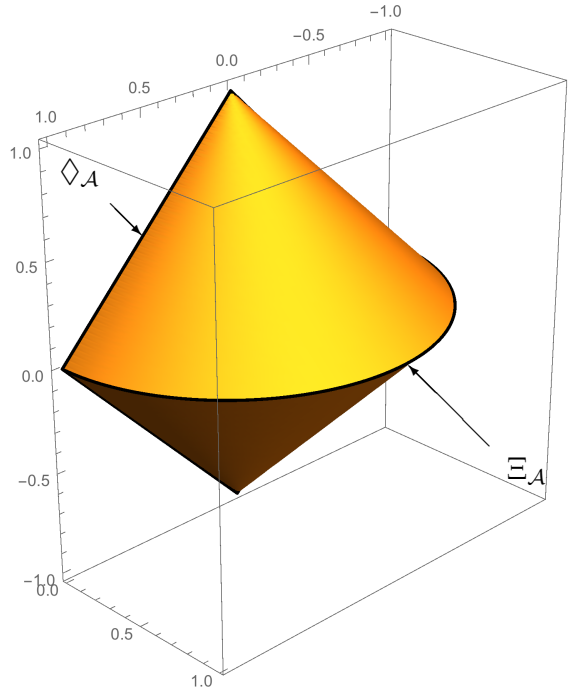

First let us review the simple example of the causal wedge in Poincar\'{e} AdS$_3$.  The relevant equations are \eqref{eq:nullgeodesics} with $z=1$.  We choose our boundary region ${\cal A}$ to be an interval of width $2a$ on a time slice:  $\mathcal{A}=\left\{(t,x)\, |\, t = 0, |x|\leq a\right\}$.  The boundary domain of dependence associated with this region is 
\begin{equation}
\domd = \left\{(t,x) \left|\, |t| \leq a-x, x \in \left[0,a\right] \right\} \right. \cup  \left\{(t,x)\left|\, |t| \leq a+x, x \in \left[-a,0\right] \right\} \right. .
\end{equation}
The future  boundary $\bcwedgef$ of the causal wedge is given by null geodesics sent from the future tip of $\domd$ at $(t,x)=\left( a,0\right)$. The result is
\begin{equation}\label{eq:futurewedgebdyAdS}
\bcwedgef = \left\{(t,x,u)\left|\, u=\sqrt{\left(a-t\right)^2-x^2}, t \in \left[0,a\right], |x|\leq a \right\} \right. .
\end{equation}
The past boundary $\bcwedgep$ is defined similarly, with $t\to -t$.  These two null surfaces  intersect at the $t=0$ slice in the bulk. The causal wedge itself is the entire bulk region sandwiched between (and including) these surfaces (the past and future Rindler horizons) and $\domd$. We show an example in figure~\ref{fig:causalwedgeAdS}.

The  past and future boundaries of the causal wedge intersect at a bulk co-dimension two surface called the  causal  information surface $\Xi_{\cal A}$  \cite{Hubeny:2012wa}.  In the previous example, this coincides with the spacelike geodesic anchored at the endpoints of $\cal A$.  For a spacetime that is only asymptotically AdS, this is not true in general. In fact, for spacetimes satisfying the null energy condition it can be shown that an extremal surface anchored on the same $\partial \cal A$ must lie outside (or on the boundary of) the causal wedge \cite{Hubeny:2012wa, Wall:2012uf, Hubeny:2013gba}.  This issue is more complicated in Lifshitz spacetime, as we will see.

\subsection{Causal wedges in Lifshitz spacetime}

Now we turn to Lifshitz spacetimes \eqref{eq:Lifshitzmetric} with $z>1$. As mentioned in section~\ref{sec:intro}, such spacetimes do not have a conformal boundary in the usual sense.  Our prescription is to cut off the spacetime at a slice of constant radius $\{u=\varepsilon\}$.  We can define a domain of dependence on this cut-off surface for any non-zero $\varepsilon$ and we make the minimal modifications necessary to all other definitions. The purpose of this section is to study the resulting causal wedge and explore what happens as we remove the cut-off.  

\begin{figure}[ht!]
\begin{center}
\includegraphics[width=0.5\textwidth]{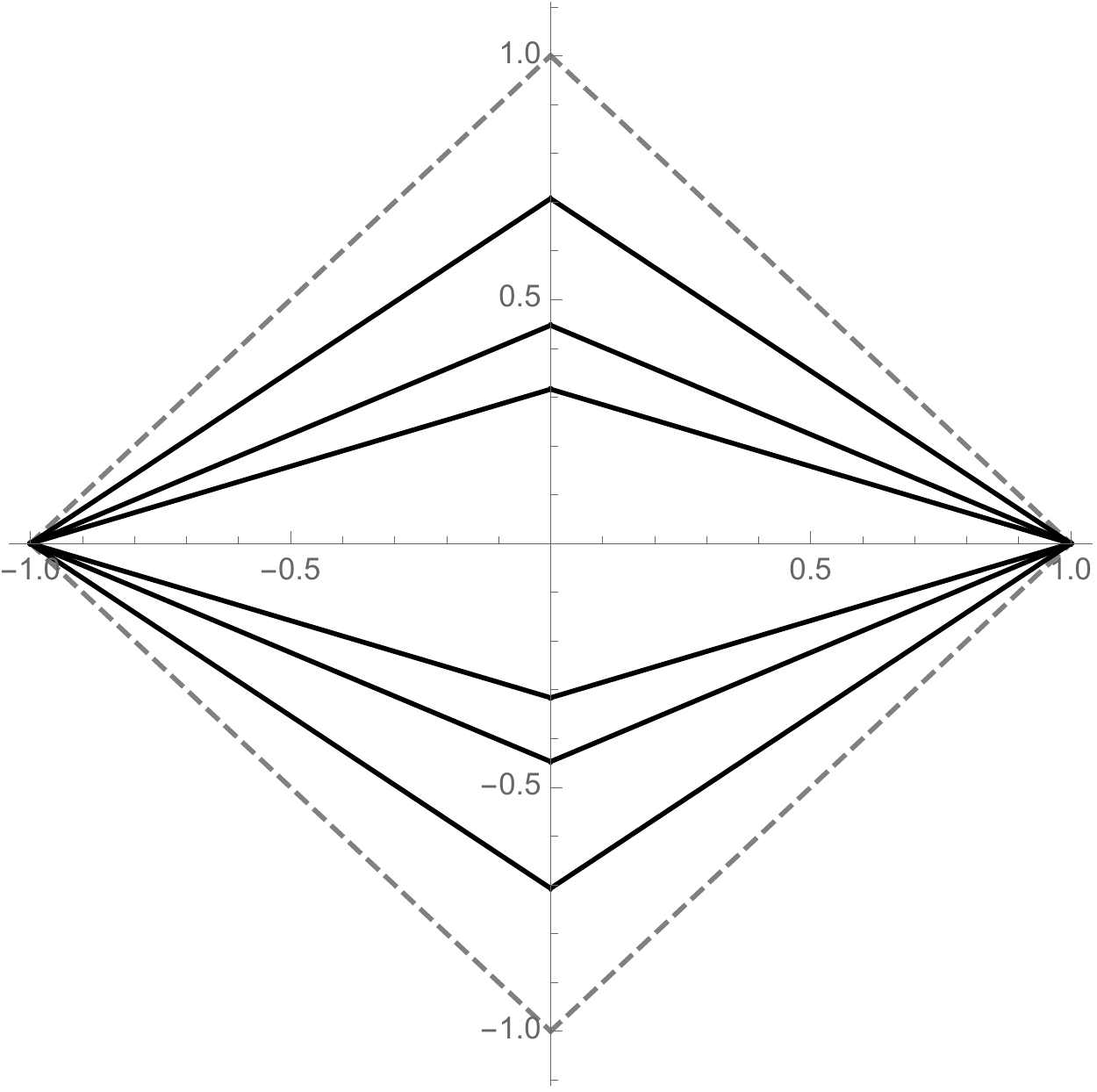}
\setlength{\unitlength}{0.1\columnwidth}
\begin{picture}(0.3,0.4)(0,0)
\put(-3.4,4.7){\makebox(0,0){$t$}}
\put(0.3,2.5){\makebox(0,0){$x$}}
\end{picture}
\end{center}
\vskip-1em
\caption{Regulated boundary domain of dependence $\domd^\varepsilon$ in Lifshitz spacetime with $z=3/2$.  We fix the  interval width $2a=2$ and plot three different values of the cut-off $\varepsilon$: $0.5$, $0.2$ and $0.1$ (from the outside to the centre) in black. We show $\domd$ for Poincar\'{e} AdS$_3$ in gray for comparison.}\label{fig:BDDshrinkage}
\end{figure}
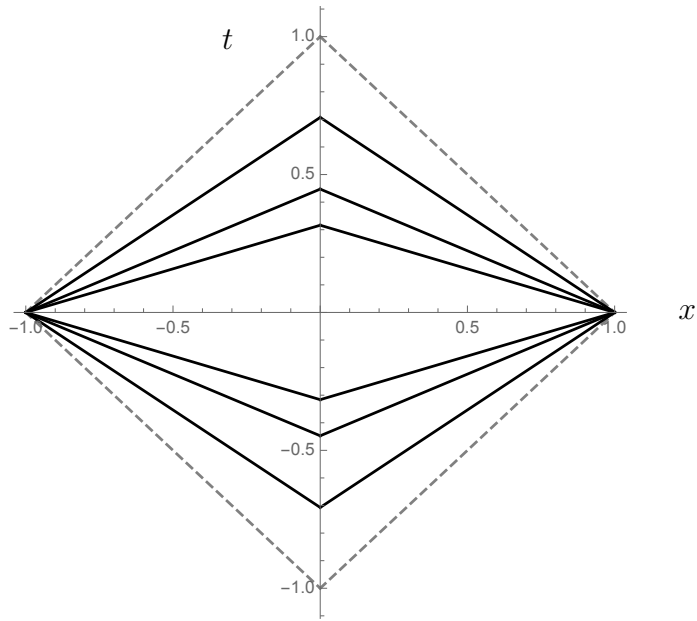

We continue to focus on an interval $\mathcal{A}=\left\{(t,x)\, |\, t = 0, |x|\leq a\right\}$, but now we define a regulated boundary domain of dependence $\domd^{\varepsilon}$ at $u=\varepsilon$ via
\begin{equation}
\domd^\varepsilon = \left\{(t,x) \left| \, |t| \leq \varepsilon^{z-1} ( a-x), x \in \left[0,a\right] \right\}\right. \cup  \left\{(t,x) \left| \, |t| \leq\varepsilon^{z-1} ( a+x), x \in \left[-a,0\right] \right\}\right. .
\end{equation}
The boundaries of this region are null geodesics of the induced metric on $\{u=\varepsilon\}$.\footnote{By definition, these null geodesics satisfy $\dot u \equiv 0$.} The future- and past-most  tips are located at $(t,x)=\left(\pm a\, \varepsilon^{z-1},0\right)$, respectively. For fixed interval width $2a$, note that $\domd^\varepsilon$ flattens as $\varepsilon$ is lowered, as demonstrated in figure~\ref{fig:BDDshrinkage}.  This is a consequence of the non-relativistic causal structure at the boundary, wherein the causal past of a point (or spatial region) includes everything in its past.

We define the bulk causal wedge $\cwedge^\varepsilon$ in the cut-off Lifshitz spacetime to be  the intersection of the causal past and the causal future of $\domd^\varepsilon $. Its boundaries in the bulk are constructed by sending null geodesics into the bulk from the boundary of $\domd^\varepsilon $.  Rewriting  \eqref{eq:nullgeodesics}, we must therefore solve
\begin{align}
\frac{dt}{du} &= \mp  \frac{u^{2(z-1)}}{\sqrt{u^{2(z-1)}-\ell^2}} , \quad t(\varepsilon) = \pm t_0\label{eq:tofueqn};
\\
\frac{dx}{du} &= \mp \frac{\sqrt{\ell^2}}{\sqrt{u^{2(z-1)}-\ell^2}}, \quad x(\varepsilon) = \mp x_0 \label{eq:xofueqn}.
\end{align}
The parameters $t_0\geq 0, x_0\geq 0$ specify the distance in $t,x$ from the center $(t,x,u)=(0,0,\varepsilon)$  of $\domd^\varepsilon $. Choosing $t_0= \varepsilon^{z-1}(a-x_0)$  and allowing $x_0\in[0,a]$ with these sign choices moves us along the boundary of $\domd^\varepsilon $.

Before solving  these equations in general, let us focus on the simplest case of purely radial geodesics ($\ell = 0$) sent from the  future and  past tips of $\domd^\varepsilon $ ($t_0 = a\, \varepsilon^{z-1}, x_0=0$). The solution to \eqref{eq:tofueqn}  is then 
\begin{equation}
t_{\pm}^{(\ell=0)}(u)=\pm\left(a\, \varepsilon^{ z-1} - \frac{u^z - \varepsilon^z}{z}\right).
\end{equation}
These geodesics intersect  the $t=0$ slice (and each other) when 
 \begin{equation}\label{eq:radialextent}
u=u_{\Xi}\equiv\varepsilon \left(1+\frac{a z}{ \varepsilon} \right)^{\frac{1}{z}}.
\end{equation}
This $u_{\Xi}$ is the radial extent of the causal wedge. As well as flattening $\domd^\varepsilon$ in time, we see that lowering  $\varepsilon$ sends this radial extent to zero, and so the degeneration of the causal wedge is very severe.  

Now we present  solutions for general values of $t_0,x_0$ and $\ell$:  
\begin{align}
t_{\pm}(u)&=\pm\left( t_0 - \frac{1}{z} \left[ u\sqrt{u^{2 (z-1)}-\ell^2 } - \varepsilon\sqrt{\varepsilon^{2 (z-1)}-\ell^2 }  \right] + \frac{\ell^2}{z\left(z-2\right)}\left[ G(u) - G(\varepsilon) \right]\right) ,
\\
x_{\pm}(u)&=\pm\left( x_0 - \frac{\sqrt{\ell^2}}{z-2}\left[ G(u) - G(\varepsilon) \right]\right) ,
\end{align}
where we have defined the function
\begin{equation}
G(v) \equiv v^{2-z} \, _2F_1\left(\frac{1}{2},\frac{z-2}{2 (z-1)};\frac{4-3 z}{2-2 z};\ell^2 v^{2-2 z}\right).
\end{equation}
The expressions for $z=2$ are
\begin{align}
t_{\pm}(u)&=\pm\left( t_0 - \frac{1}{2} \left[ u\sqrt{u^{2}-\ell^2} - \varepsilon\sqrt{\varepsilon^{2}-\ell^2 }  \right] - \frac{\ell^2}{2}\log\left[ \frac{u+\sqrt{u^{2}-\ell^2} }{\varepsilon+\sqrt{\varepsilon^{2}-\ell^2} } \right]\right) ,
\\
x_{\pm}(u)&=\pm\left( x_0 + \sqrt{\ell^2}\log\left[ \frac{u+\sqrt{u^{2}-\ell^2} }{\varepsilon+\sqrt{\varepsilon^{2}-\ell^2} } \right]\right) .
\end{align}
We will see that geodesics with $t_{+}(u)$ generate $\bcwedgef$ and geodesics with $x_{+}(u)$ generate the half of $\partial\left(\cwedge\right)$  with $x\in\left[0,a\right]$. (The minus signs refer to the appropriate opposites.)

Before plotting these curves to find  the boundary of the causal wedge, let us highlight a key difference from AdS.  Suppose we send null geodesics just from the future- and past-most tips of $\domd^\varepsilon $.  This is what one would do for an asymptotically  AdS spacetime.    However, our  discussion in section~\ref{sec:conventions} implies that, for a given choice of $\varepsilon$, null geodesics with $|\ell|> \varepsilon^{z-1}$ are classically forbidden and therefore cannot be included.   In the AdS case this bound is $|\ell|=1$ and such geodesics sent from the tips run along the boundary of $\domd$. This is not the case for Lifshitz spacetimes: geodesics with the maximum $|\ell| = |\ell_{\star}| \equiv  \varepsilon^{z-1}$ sent from the tips  of $\domd^\varepsilon $ bend away from the cut-off surface into the bulk.\footnote{Such geodesics have $\dot u = 0$, $\ddot u >0$ at $u=\varepsilon$, unlike those forming $\domd^\varepsilon $, which have $\dot u \equiv 0$.}  This leads us to  suspect that the surfaces built from null geodesics sent from these tips will not close at the edges at $\{u=\varepsilon\}$.  

One could be concerned at this stage that the causal past or future of these tips is ill-defined.  However, the allowed geodesics sent from these tips in the cut-off Lifshitz spacetime do indeed  form (the curved surfaces of)  half-cones \emph{locally}.   Unlike the AdS case, this is not true away from the tips.  In order to build the full boundaries of the causal wedge we therefore need to include the null geodesics from the rest of the boundary of $\domd^\varepsilon $.  Specifically, the causal wedge boundary is built from two types of null geodesic:
\begin{description}
\item{Type I:} Null geodesics sent from the future- and past-most tips of $\domd^\varepsilon $ with $0 \leq |\ell| \leq |\ell_{\star}|$.
\item{Type II:} Null geodesics sent from other points on the  boundary of $\domd^\varepsilon $ with $|\ell| = |\ell_{\star}|$.
\end{description}
(Again, in the AdS case the latter type run along the boundary of $\domd$.)

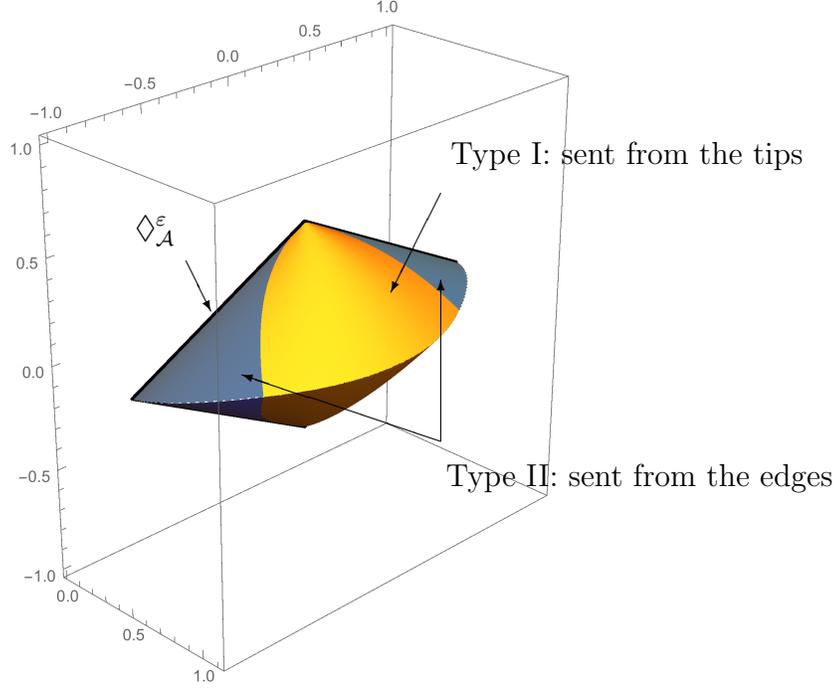
\begin{figure}[h!]
\begin{center}
\hskip.75em
\includegraphics[width=0.45\textwidth]{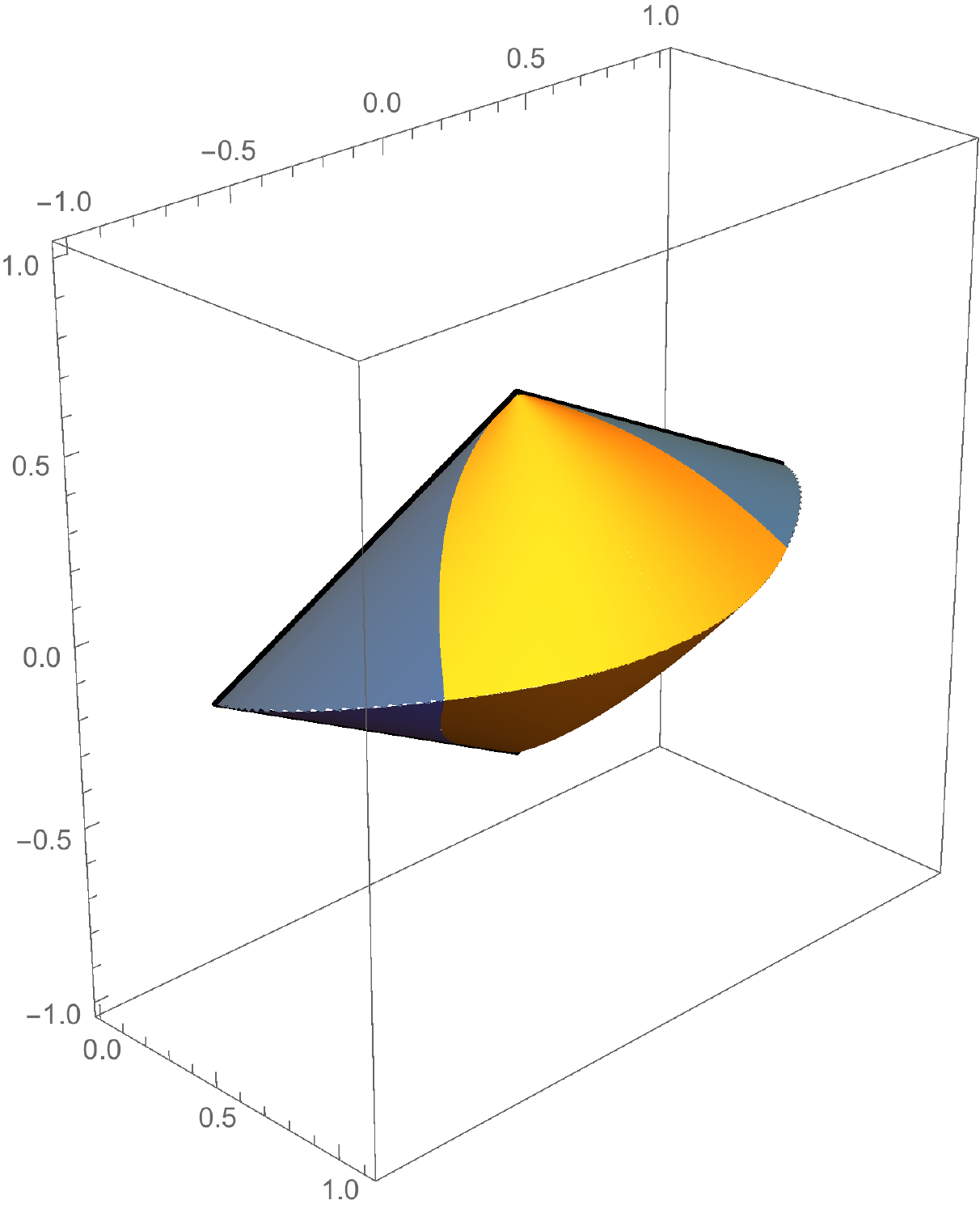}
\setlength{\unitlength}{0.1\columnwidth}
\begin{picture}(0.3,0.4)(0,0)
\put(-3.5,3.7){\makebox(0,0){$\domd^\varepsilon $}}
\put(-3.25,3.45){\vector(1,-2){0.2}}
\put(0.3,4.3){\makebox(0,0){Type I: sent from the tips}}
\put(-1.2,4){\vector(-1,-2){0.4}}
\put(0.4,1.7){\makebox(0,0){Type II: sent from the edges}}
\put(-1.2,2){\vector(0,1){1.3}}
\put(-1.2,2){\vector(-3,1){1.6}}
\end{picture}
\end{center}
\caption{The causal wedge $\cwedge^\varepsilon$ in  a cut-off Lifshitz spacetime with $z=2$. We fix the interval width $2a=2$ and also the cut-off $\varepsilon=0.5$. The central yellow section is built from light rays sent from the future- and past-most tips of the regulated boundary domain of dependence $\domd^\varepsilon$ (Type I), whereas the outer blue sections are built from light rays  sent from its edges with $|\ell|=|\ell_\star|$ (Type II).}\label{fig:causalwedge}
\end{figure}

Finally we are ready to  present the  causal wedge  $\cwedge^\varepsilon$ in  the cut-off Lifshitz spacetime.  In figure~\ref{fig:causalwedge} we plot an example with $z=2$.  This looks qualitatively similar to the wedge for Poincar\'{e} AdS$_3$ presented in figure~\ref{fig:causalwedgeAdS}, besides its  height being rescaled by a factor of $\varepsilon^{z-1}$.  However, it is clear that Type II geodesics are required in order to form a closed co-dimension zero wedge  of the bulk.  

For different values of $z$, $a$ and $\varepsilon$ we find qualitatively similar wedges.  To visualize the quantitative dependence on these parameters it is useful to focus on the causal information surface  $\Xi_{\cal A}$, which is a curve in three bulk dimensions.  Whilst we have been unable to determine this curve analytically, it is straightforward to find it numerically. For fixed $z$, in  figure~\ref{fig:CHI_z2}  we see how the causal wedge degenerates as we lower the cut-off $\varepsilon$.  Note that the radial extents match the formula given in \eqref{eq:radialextent}. For fixed $\varepsilon$, in figure~\ref{fig:CHI_varyz} we observe the same effect as $z$ is increased.
\begin{figure}[h!]
\vskip1em
\begin{center}
\includegraphics[width=0.5\textwidth]{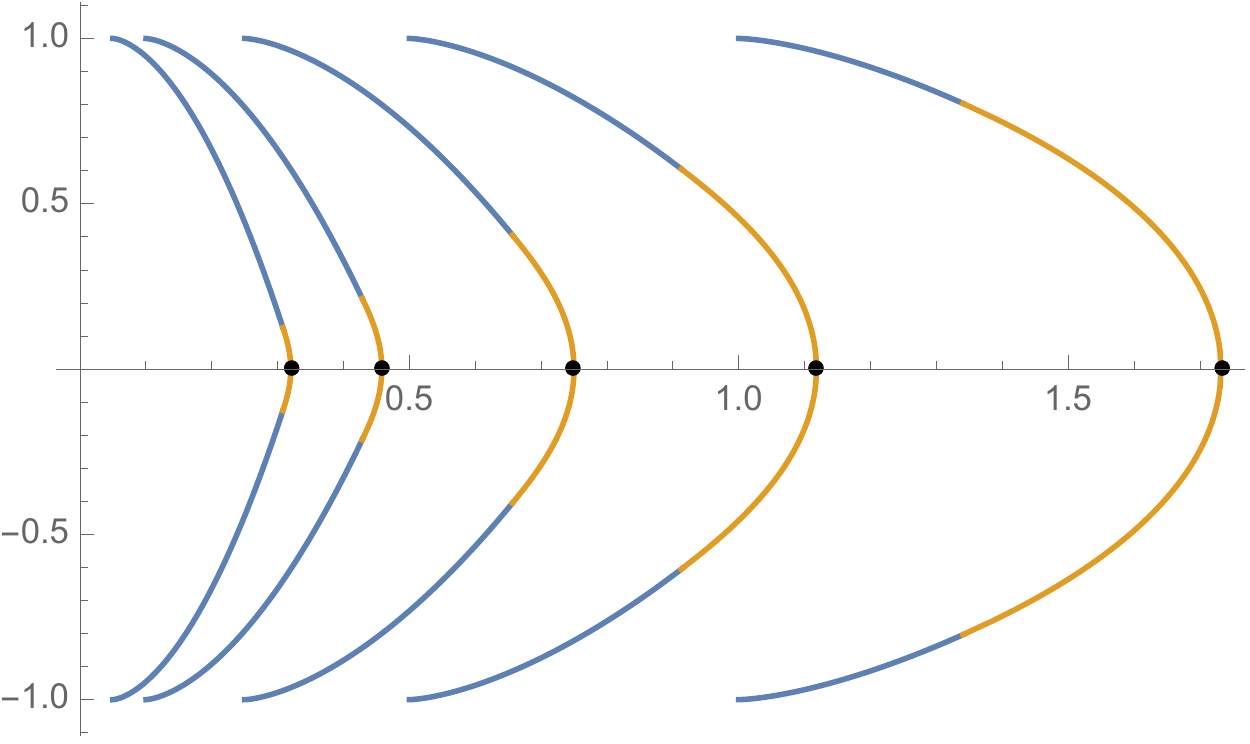}
\setlength{\unitlength}{0.1\columnwidth}
\begin{picture}(0.3,0.4)(0,0)
\put(-5.5,3.){\makebox(0,0){$x$}}
\put(0.2,1.5){\makebox(0,0){$u$}}
\end{picture}
\end{center}
\vskip-1em
\caption{Examples of the causal information surface $\Xi_{\cal A}$ in Lifshitz spacetime with $z=2$. We fix the interval width $2a=2$ and find the surface for five different values of the cut-off $\varepsilon$: $0.05$, $0.1$, $0.25$, $0.5$ and $1$,  from left to right. For each curve, the central yellow section is built from Type I light rays, whereas the outer blue sections are built from Type II light rays, as in figure~\ref{fig:causalwedge}.  Recall that $|\ell_\star|=\varepsilon^{z-1}$. Black dots mark the radial extent $u_{\Xi}$ in each case via \eqref{eq:radialextent}.}\label{fig:CHI_z2}
\end{figure}
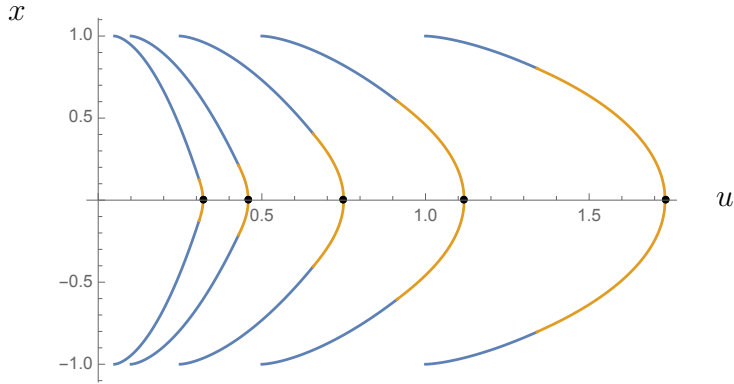
\begin{figure}[h!]
\vskip1em
\begin{center}
\includegraphics[width=0.5\textwidth]{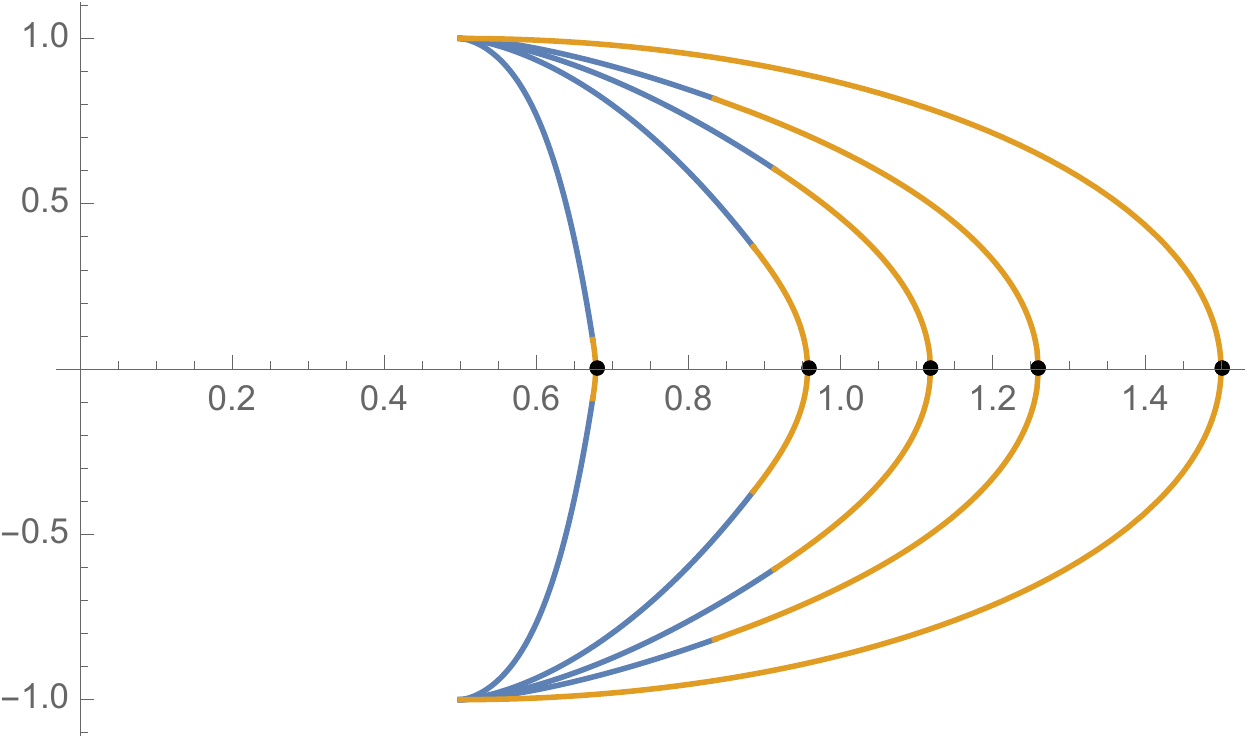}
\setlength{\unitlength}{0.1\columnwidth}
\begin{picture}(0.3,0.4)(0,0)
\put(-5.5,3.){\makebox(0,0){$x$}}
\put(0.2,1.5){\makebox(0,0){$u$}}
\end{picture}
\end{center}
\vskip-1em
\caption{Examples of the causal information surface $\Xi_{\cal A}$ at various $z$. We fix the interval width $2a=2$ and also the cut-off $\varepsilon=0.5$.  From right to left: $z=1$, $3/2$, $2$, $3$ and $10$.}\label{fig:CHI_varyz}
\end{figure}

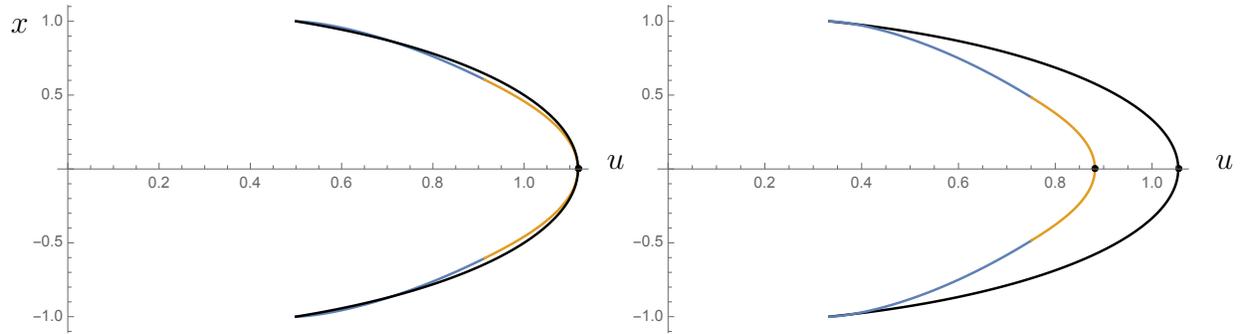
\begin{figure}[h!]
\begin{center}
\hskip0.75em
\includegraphics[width=0.45\textwidth]{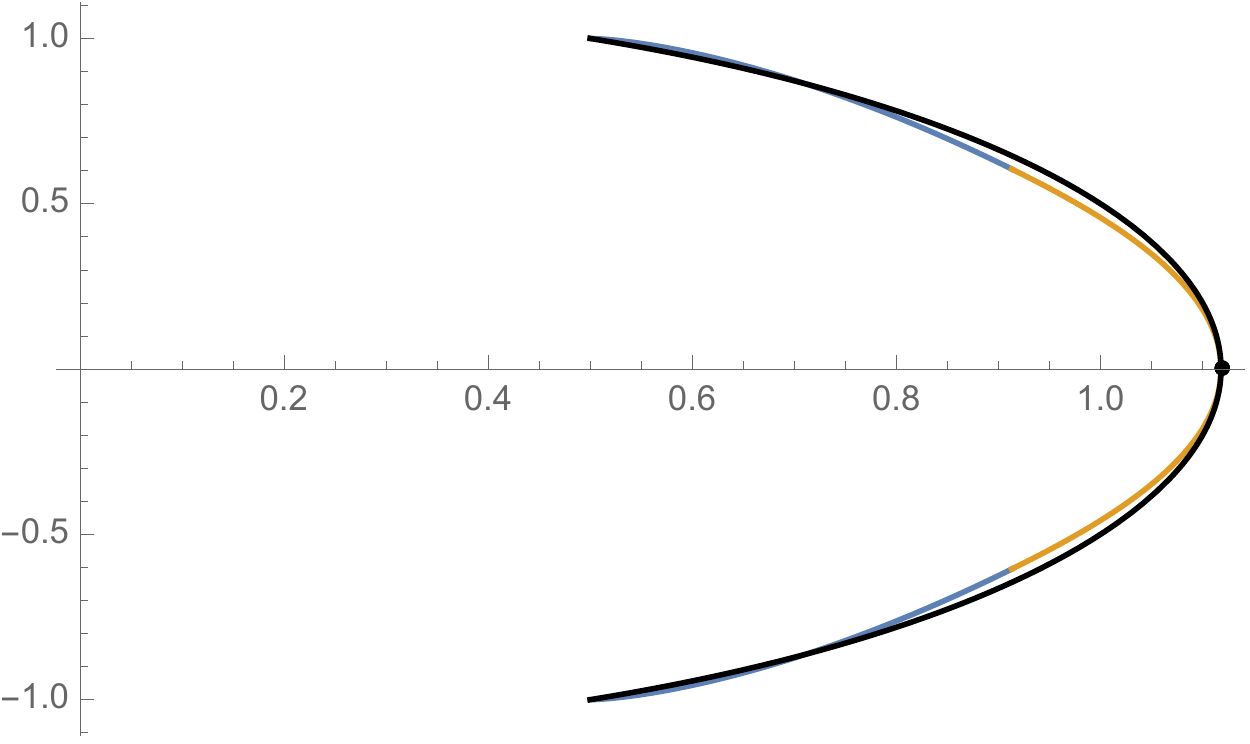}
\hskip1em
\includegraphics[width=0.45\textwidth]{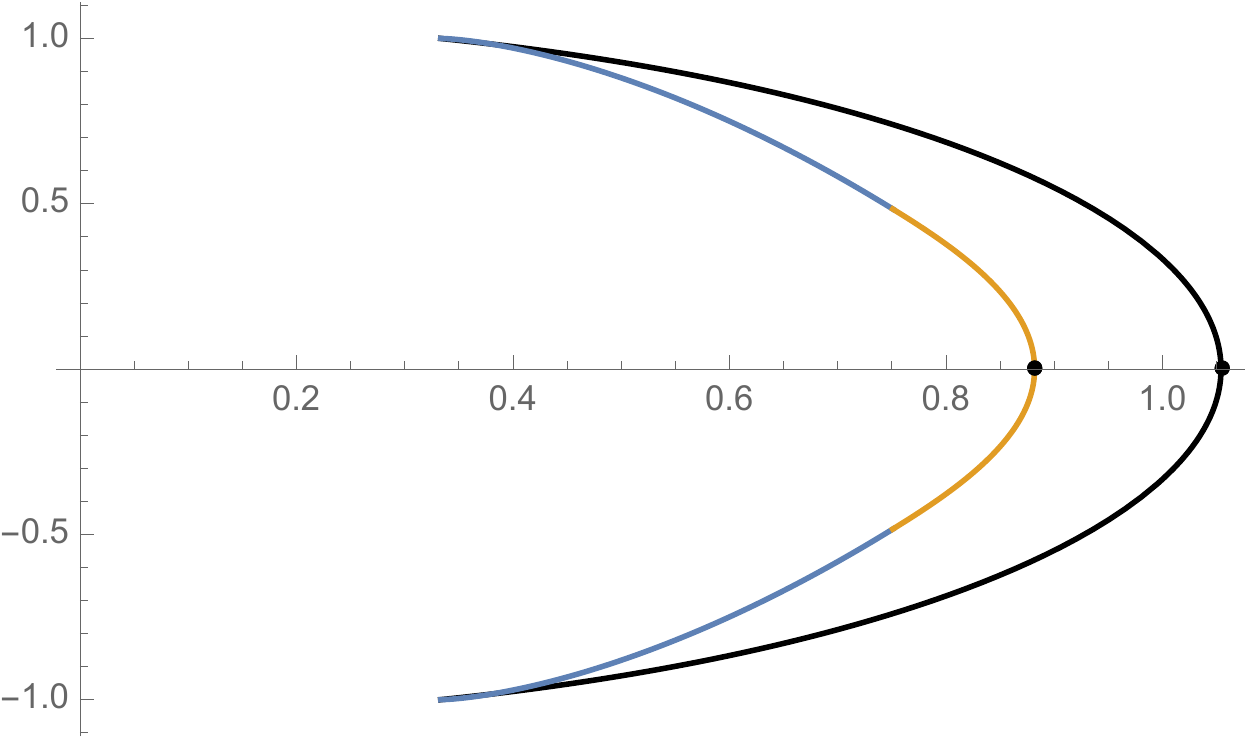}
\setlength{\unitlength}{0.1\columnwidth}
\begin{picture}(0.3,0.4)(0,0)
\put(-9.6,2.5){\makebox(0,0){$x$}}
\put(-4.8,1.4){\makebox(0,0){$u$}}
\put(0.1,1.4){\makebox(0,0){$u$}}
\end{picture}
\end{center}
\caption{Causal information curves $\Xi_{\cal A}$ for  $z=2$ and the spacelike geodesic (black curves) with the same endpoints.  We present one example for which $u_\star=u_\Xi$ ($\varepsilon=1/2$, left)  and one for which $u_\star>u_\Xi$ ($\varepsilon=1/3$, right).  Black dots mark $u_\Xi$ using \eqref{eq:radialextent} and  $u_\star$ using \eqref{eq:newustar}.}\label{fig:CISneqgeodesic}
\end{figure}

As mentioned in section~\ref{sec:AdSwedges}, the causal information surface  in Poincar\'{e} AdS$_3$ coincides with the spacelike geodesic anchored at the same endpoints.   A natural question is: does this also hold  for Lifshitz spacetime if we define a spacelike geodesic anchored on the cut-off surface?  The formula for the latter is known for all $z$ analytically because it lies on a time slice (i.e.\ it has $E=0$): 
\begin{equation}\label{eq:semicircle}
x_{\pm}(u) =\pm\, \frac{\sqrt{1 - P^2 u^2}}{P}.
\end{equation}
Its momentum $P$ (and hence turning point $u_{\star}$) is determined by demanding that its endpoints are anchored at the interval endpoints $x = \pm a$ on the cut-off surface at $u = \varepsilon$:
\begin{equation}\label{eq:newustar}
P = \frac{1}{\sqrt{a^2 + \varepsilon^2}} \quad \Rightarrow \quad u_{\star} = \sqrt{a^2 +  \varepsilon^2} .
\end{equation}
In figure~\ref{fig:CISneqgeodesic} we demonstrate that the answer to the above question is negative. In general the two curves have only the endpoints in common. 

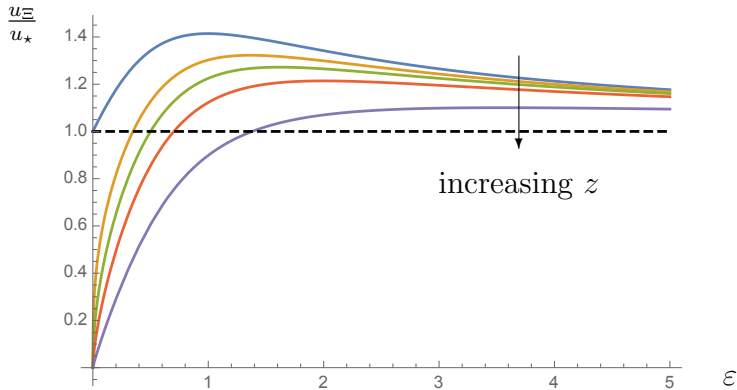
\begin{figure}[h!]
\vskip1em
\begin{center}
\includegraphics[width=0.5\textwidth]{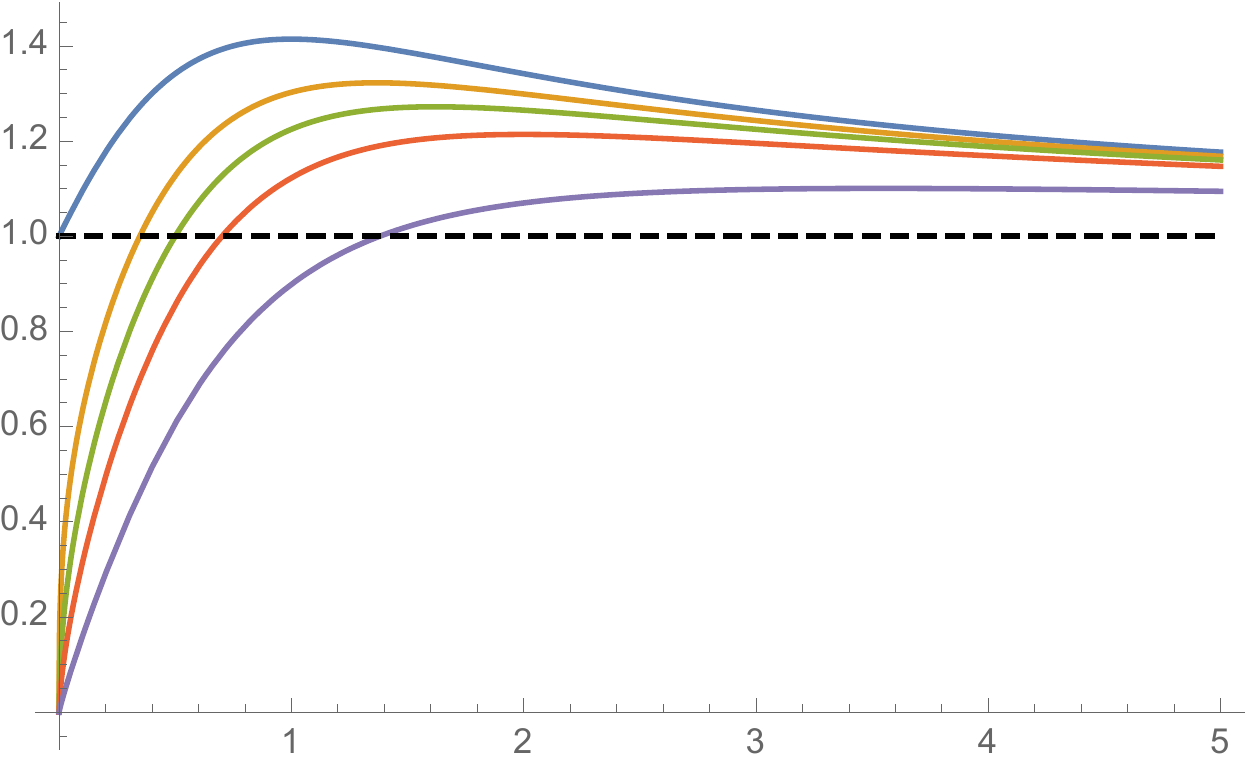}
\setlength{\unitlength}{0.1\columnwidth}
\begin{picture}(0.3,0.4)(0,0)
\put(-5.5,3.){\makebox(0,0){$\frac{u_\Xi}{u_\star}$}}
\put(0.2,0.15){\makebox(0,0){$\varepsilon$}}
\put(-1.5,1.7){\makebox(0,0){increasing $z$}}
\put(-1.5,2.75){\vector(0,-1){0.75}}
\end{picture}
\end{center}
\vskip-1em
\caption{A comparison between the radial extent $u_{\Xi}$ of the causal wedge and the  turning point $u_{\star}$ of the spacelike geodesic as a function of the cut-off $\varepsilon$ for different values of $z$: 1 (blue), 3/2 (yellow), 2 (green), 3 (red)  and 10 (purple) at fixed $a=1$.  The ratio tends to unity as $\varepsilon\to\infty$ for fixed $z$ (and also as $z\to\infty$ for fixed $\varepsilon$).}\label{fig:uXivsustar}
\end{figure}

However, for sufficiently large $\varepsilon$ the  turning point $u_{\star}$ of the spacelike geodesic is \emph{less than} the radial extent $u_{\Xi}$.  Said differently, the causal wedge can reach further into the bulk than the spacelike geodesic anchored at the same points.  This is demonstrated in figure~\ref{fig:uXivsustar}.  This appears to be at odds with the general result of \cite{Hubeny:2012wa, Wall:2012uf, Hubeny:2013gba} for an asymptotically AdS spacetime mentioned earlier. How can we resolve this? 

The key point is that the proof by contradiction presented in those papers does not hold when we introduce a cut-off.  Unlike the boundary, the cut-off surface can be reached by a light ray from a bulk point on the causal information surface in finite affine parameter.  As a result, the expansion of $\Xi_{\cal A}$ can be non-negative without leading to  caustics.%
\footnote{One can demonstrate this by manipulating Raychaudhuri's equation --- see section 9.2 of Wald \cite{Wald:1984rg}, for example.} %
  We introduce a cut-off in order to define the causal wedge in Lifshitz spacetime, which is why we see $u_\Xi / u_\star$ exceed unity as we increase $\varepsilon$.  Of course, our primary interest is in removing the cut-off and in the  regime of small $\varepsilon$ we recover $u_\Xi < u_\star$.

The area of the causal  information surface in Planck units was dubbed causal holographic information $\chi_{\cal A}$ in \cite{Hubeny:2012wa}. Whilst there is no clear understanding of $\chi_{\cal A}$ from field theory as yet,\footnote{See however the interesting recent proposal of \cite{Kelly:2013aja} and earlier attempts by \cite{Freivogel:2013zta}.} it was conjectured in \cite{Hubeny:2012wa} that this should provide an upper bound on the holographic entanglement entropy associated with the same region $\cal A$.  When $\Xi_{\cal A}$ and the spacelike geodesic lie on a constant-time slice, clearly we should have  $\chi_{\cal A}\geq S_{\cal A}$ because the geodesic is the minimal length curve anchored at $\partial {\cal A}$. The closer a curve hugs the boundary, the greater its length, so this implies $u_\star\geq u_\Xi$.  But we have just seen that this latter condition is not true in Lifshitz spacetime for large $\varepsilon$, so this is a further warning that care must be taken when applying AdS intuition  to Lifshitz.

Causal wedges  have  been constructed in asymptotically AdS solutions: for example, in Schwarzschild-AdS black holes \cite{Hubeny:2013gba} and in horizonless  scalar solitons \cite{Gentle:2013fma}.  Generically, the causal wedge can have non-trivial topology and so $\Xi_{\cal A}$  will be a proper subset of the $t=0$ surface  consisting of a number of disconnected regions \cite{Hubeny:2013gba}. Both of these examples exhibit this feature, even though the latter spacetimes are causally trivial.

Note  that we do not see holes in our cut-off wedges in Lif$_3$.  As argued in \cite{Hubeny:2013gba}, the causal wedge cannot have non-trivial topology in three-dimensional bulk spacetimes.  The higher dimensional examples mentioned above are  spherically symmetric spacetimes, but this symmetry is not a necessary condition for holes.  It is therefore reasonable to expect that causal wedge holes can exist in asymptotically Lifshitz solutions in higher dimensions.

In conclusion, sensible causal wedges  can be defined in Lifshitz spacetime if we introduce a radial cut-off. These wedges  look very similar to those found in Poincar\'{e} AdS$_3$. Degeneration of the causal wedge as we remove this cut-off is inevitable, in line with a non-relativistic boundary theory.

The cut-off imposed here serves a qualitatively different purpose  to the one used in section~\ref{sec:differentialentropy}.  There, a radial cut-off was introduced simply to regulate the length of a bulk spacelike geodesic.  It dropped out of the construction in the end.  In contrast, here a radial cut-off was required in order to define the causal wedge at all.

\section{The entanglement wedge}%
\label{sec:entanglementwedge}

In this section, we construct the entanglement wedge considered in \cite{Czech:2012bh, Headrick:2014cta}.  The entanglement wedge is one conjecture for the most natural bulk region corresponding to the boundary reduced density matrix. 

Consider a boundary region ${\cal A}$ that is a subset of a boundary Cauchy slice.  If the  slice in question aligns with a bulk Killing vector rendering the spacetime static, we follow the Ryu-Takayanagi procedure to compute the holographic entanglement entropy for ${\cal A}$.  Otherwise, we use the covariant procedure proposed in \cite{Hubeny:2007xt}.  In both cases, the holographic entanglement entropy is the (regulated) area of the extremal co-dimension two surface ${\cal E}_{\cal A}$ anchored at the boundary of (and homologous to) the region ${\cal A}$.

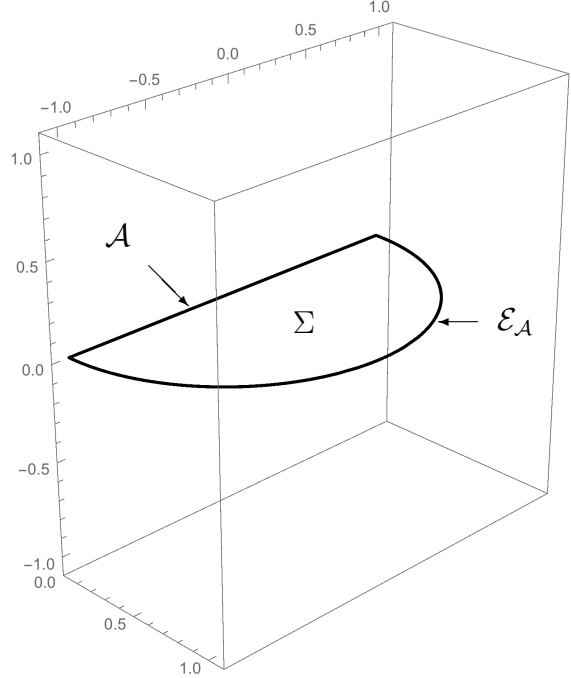
\begin{figure}[h!]
\begin{center}
\hskip.75em
\includegraphics[width=0.45\textwidth]{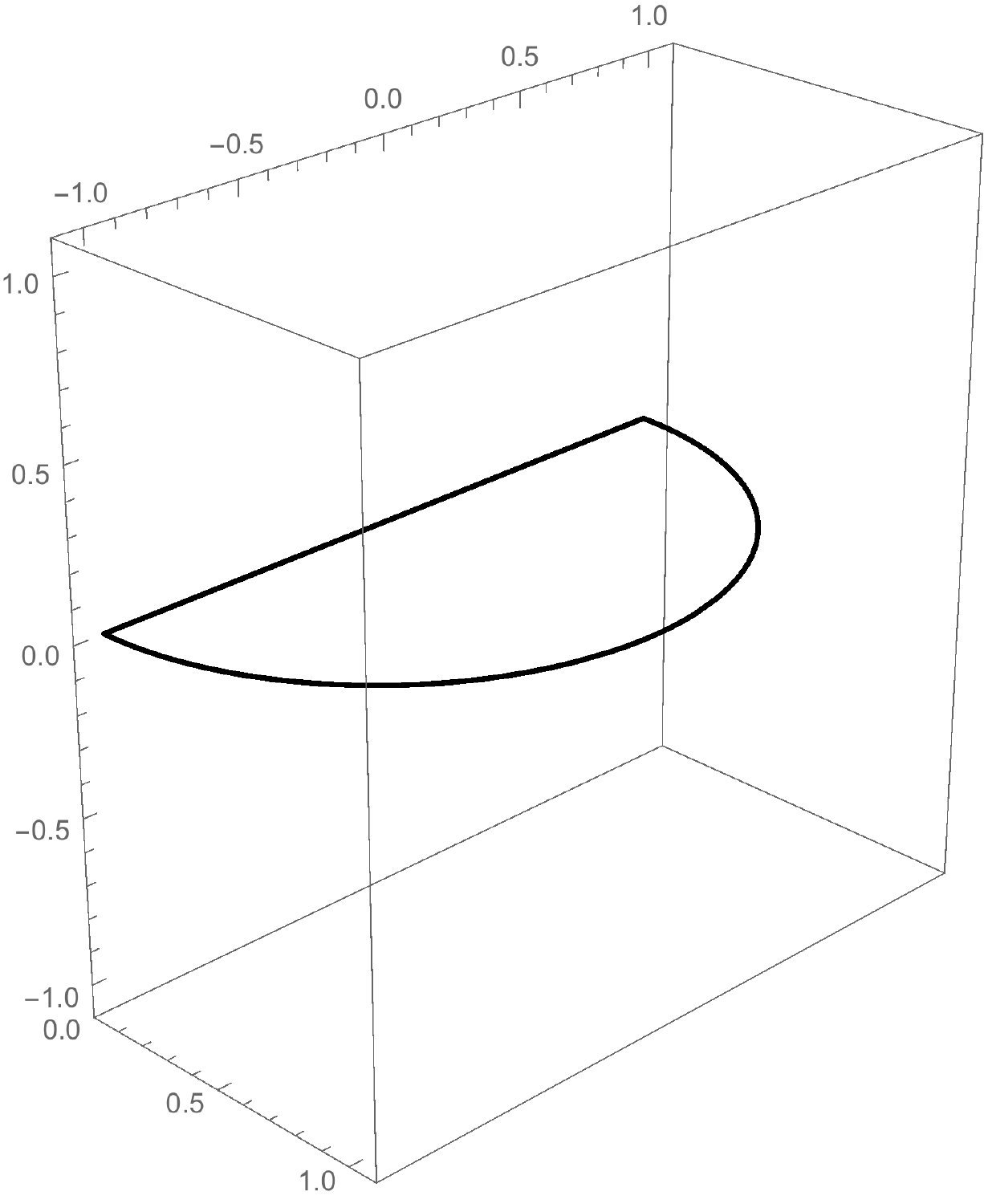}
\setlength{\unitlength}{0.1\columnwidth}
\begin{picture}(0.3,0.4)(0,0)
\put(-3.8,3.6){\makebox(0,0){${\cal A}$}}
\put(-3.55,3.35){\vector(1,-1){0.32}}
\put(-2.3,2.9){\makebox(0,0){$\Sigma$}}
\put(-0.6,2.9){\makebox(0,0){${\cal E}_{\cal A}$}}
\put(-0.9,2.9){\vector(-1,0){0.32}}
\end{picture}
\end{center}
\caption{This figure shows the region $\Sigma$ enclosed by the boundary subregion ${\cal A}$ and the associated extremal  curve ${\cal E}_{\cal A}$.}
\label{fig:Sigmaplot}
\end{figure}

Together, ${\cal E}_{\cal A}$ and ${\cal A}$ enclose a spacelike co-dimension one region, which we call $\Sigma$.  This satisfies $\partial \Sigma= {\cal A}\cup{\cal E}_{\cal A}$ and an example is illustrated in figure~\ref{fig:Sigmaplot}.  The entanglement wedge, denoted ${\cal W}_{\cal A}$, is defined to be the causal development of $\Sigma$.

In order to find the causal development of $\Sigma$, we begin by considering light-sheets emanating from ${\cal E}_{\cal A}$.   As a co-dimension two spacelike surface, we can use ${\cal E}_{\cal A}$ to define a pair of light-sheets following the construction in \cite{Bousso:2002ju}.  We build these light-sheets by finding the null geodesics at each point on ${\cal E}_{\cal A}$ that are orthogonal to ${\cal E}_{\cal A}$.  This produces four sets of null geodesics: one pair heading towards the boundary and one pair heading towards the bulk, where one of each pair is headed towards the future and one towards the past.  Additionally imposing that the expansion $\theta$ of the null congruence associated with this set of orthogonal geodesics satisfy $\theta\leq 0$ picks out two sets of null geodesics; these two sets of null geodesics comprise the light-sheets associated with the surface ${\cal E}_{\cal A}$. 

In the AdS case, these light-sheets from ${\cal E}_{\cal A}$ are sufficient: the entanglement wedge is the bulk region bounded on one side by the light-sheets built from ${\cal E}_{\cal A}$, and on the other by the  boundary domain of dependence $\domd$ of the region ${\cal A}$.  However, for Lifshitz spacetimes, the light-sheets emanating from ${\cal E}_{\cal A}$ do not form a closed surface; they do not reach $u=0$.  We must also include light-sheets coming from into the bulk from the boundary region ${\cal A}$ in order to find the causal development of the region $\Sigma$.

For our explicit constructions, we will focus on three-dimensional bulk spacetimes; thus, the extremal surface ${\cal E}_{\cal A}$ will be just the spacelike geodesic whose end points coincide with the endpoints of the interval ${\cal A}$ on the boundary.  Thus they will satisfy the geodesic equation \eqref{eq:gengeodesics} with $\kappa =1$.  

\subsection{Entanglement wedge in  AdS$_3$}
We begin by studying the entanglement wedge in vacuum AdS$_3$ for a region on a constant time slice.  As before, we choose ${\cal A}=\left\{(t,x)\, |\,  t=0,|x|\leq a \right\}$.  The spacelike geodesic homologous to ${\cal A}$ will also reside on a constant time slice and thus have $E=0$ in the geodesic equation (\ref{eq:gengeodesics}).  It is described by the equation (\ref{eq:semicircle}) with $P=1/a$. Note this equation is also valid for spacelike geodesics on a constant time slice of a Lifshitz spacetime with any value of $z$, in addition to the AdS or $z=1$ case.

Next, we need to find the null geodesics orthogonal to the spacelike geodesic in (\ref{eq:semicircle}).  For simplicity we will focus on the $x_-$ region of the spacelike geodesic; those for the $x_+$ region can be found similarly.  Writing the tangent to the spacelike geodesic as $T^\mu$, and the null geodesic as $n^\mu$, the equations
\begin{align}
n^\mu n_\mu = 0 , \label{eq:nulllightsheet}
\\
T^\mu n_\mu = 0, \label{eq:ortholightsheet}
\end{align}
enforce nullness and orthogonality to the spacelike geodesic for the null rays.  Taking the derivative of (\ref{eq:semicircle}) to find the ratio $T^x/T^u$, the second condition is solved by
\begin{equation}\label{eq:orthoallz}
\frac{n^x}{n^u} = -\frac{a}{u_s} \sqrt{1-u_s^2/a^2},
\end{equation}
where $u_s$ labels the point of intersection on the spacelike geodesic, and we have chosen the sign to select the light rays going towards the boundary.   The nullness condition on the vector $n^\mu$ becomes
\begin{equation}\label{eq:nullz=2}
\frac{n^t}{n^u} = \pm \frac{a}{u_s}.
\end{equation}
Note that $n^\mu$ here tells us the null vector for a null geodesic normal to our spacelike geodesic, at a particular location labelled by the radius $u_s$ of the intersection on that spacelike geodesic.  The minus sign indicates light rays heading towards the boundary in the future, while the plus sign indicates those from the boundary in the past. If we wish to now follow this null ray as it continues towards the boundary, we should compute the conserved quantity $\ell$ in (\ref{eq:nullgeodesics}) and then solve the null geodesic equation, with the constraint that the null ray pass through the spacelike geodesic at radius $u=u_s$.  For $\ell$ we find
\begin{equation}\label{eq:ellnormal}
\ell = \frac{n^x}{n^t} = \pm\sqrt{1-u_s^2/a^2}.
\end{equation}
So the conserved quantity $\ell$ for the null geodesic depends on the radial location $u_s$ of its intersection with the spacelike geodesic.  The null geodesics we are interested in should thus satisfy (\ref{eq:nullgeodesics}) with $\ell$ as in (\ref{eq:ellnormal}), and they should go through the point $u=u_s, \, t=0, \, x = -a\sqrt{1-u_s^2/a^2}$.  These null geodesics satisfy
\begin{align}
t &= \pm a \frac{u-u_s}{u_s},
\\
x &= -a \frac{u}{u_s}\sqrt{1-u_s^2/a^2}.
\end{align}
Each future-directed light ray here passes through the extremal surface at $u=u_s$, and then heads in a straight line towards the boundary point $u=0,\,t=a,\,x=0$. Thus the future-directed rays form one quarter of a cone whose base is the semicircle of the spacelike geodesic.  Including the rays that intersect with the $x_+$ portion of the semicircle produces a half-cone whose flat edge lies along the boundary of the spacetime.  The past directed rays here (plus those from $x_+$) produce an equivalent past half-cone.
These two half cones, plus the spacetime boundary, enclose the entanglement wedge of the boundary  region ${\cal A}$ (see figure~\ref{fig:AdSentwedge} in the following subsection).  In vacuum AdS this region coincides with the causal wedge studied in section~\ref{sec:causalwedges}.

One important feature of the cone shape here is that null rays leaving the spacelike geodesic from very close to the boundary stay in the neighbourhood of the boundary causal diamond. That is, the light-sheets leaving the bulk holographic entanglement surface smoothly connect to the boundary casual diamond.  This is not an accident; in fact, this feature was a requirement for the covariant HRT proposal \cite{Hubeny:2007xt}.  The important feature here is that light-sheets limit the amount of entropy that can flow through them; the light-sheet we have just found for the region ${\cal A}$ is thus a reasonable conjecture for the bound of the bulk spacetime region reconstructible from only boundary information within ${\cal A}$.

\subsection{Entanglement wedge in Lif$_3$}
We now move on to study the entanglement wedge for arbitrary $z$, staying in  three dimensions for simplicity.  We also choose the constant time slice $t=0$, both to avoid questions as to the boundary physical meaning of other slices as well as for simplicity.  Lastly, we maintain the same boundary region ${\cal A}$ as in the previous section.

 Again, the extremal surface is a semicircle given by (\ref{eq:semicircle}) with $P=1/a$.  Similarly, regardless of the value of $z$, the orthogonality condition is still (\ref{eq:orthoallz}).  However the nullness condition (\ref{eq:nulllightsheet}) now becomes
\begin{equation}
\frac{n^t}{n^u} = \pm a u_s^{z-2} ,
\end{equation}
and similarly the conserved momentum $\ell$ becomes
\begin{equation}\label{eq:ellnormalz}
\ell = \frac{n^x}{n^t} u_s^{2z-2} = \pm u_s^{z-1}\sqrt{1-u_s^2/a^2}.
\end{equation}
Even before we explicitly solve for the light paths, we can already see a concern here.  From the null geodesic equations (\ref{eq:nullgeodesics}), we see for $z>1$, any geodesic with nonzero $\ell$ will have a minimum radius of $\ell^{1/(z-1)}$.  There are only three future-directed rays in the light-sheet that have $\ell=0$: one leaving from $u_s=a,\, x=0$ and two from $u_s=0, \, x=\pm a$. Thus we might worry that the entanglement `wedge' in Lifshitz does not close; that is, the light-sheet itself may not reach the boundary, so there is no finite subregion of the bulk bounded by the past and future light-sheets of ${\cal E}_{\cal A}$ alone.  This is exactly the behaviour we observe below.

In order to present explicit solutions, we now specify to the case $z=2$. Next, we solve (\ref{eq:nullgeodesics}) with this $\ell$ as in (\ref{eq:ellnormalz}) and insist the geodesics pass through $u=u_s,\, t=0,\, x=-a\sqrt{1-u_s^2/a^2}$. The future light sheet is then described by the paths
\begin{align}\label{eq:xlightsheet}
x &= \sqrt{1-u_s^2/a^2}\left( \pm u_s\log \left[\frac{au-\sqrt{a^2u^2-a^2 u_s^2+u_s^4}}{a u_s \mp u_s^2}\right]-a \right) ,
\\\label{eq:tlightsheet}
t & = 
\frac{1}{2a^2}\left(
u_s^3 a \mp au \sqrt{a^2u^2-a^2u_s^2+u_s^4} \pm a^2u_s^2 \log \left[ \frac{au-\sqrt{a^2u^2-a^2u_s^2+u_s^4}}{au_s \mp u_s^2} \right]\right.
\\\nonumber
& \hspace{1.6cm}\left.+u_s^4 \log\left[ \frac{au\pm\sqrt{a^2u^2-a^2u_s^2+u_s^4}}{au_s+u_s^2} \right]
\right).
\end{align}
Here the top signs indicate paths before they reach their turning points, and the bottom signs indicate rays after the turning points, continuing in towards the bulk.

Before we present a plot of the light-sheets, we note that the (future) light-sheets only continue as long as their expansion $\theta\leq0$; that is, as soon as the light rays form a caustic, the sheet stops \cite{Bousso:2002ju}.  Caustics occur when neighbouring rays intersect (that is, when $\theta=-\infty$). Beyond a caustic, $\theta$ becomes positive; the light rays have crossed and are now going away from each other.  We can calculate the location of the caustics by finding intersections between rays starting at $u_s$ and $u_s+\delta$, then taking $\delta$ to zero.  Equivalently, we fix $u$, take the derivative of both (\ref{eq:xlightsheet}) and (\ref{eq:tlightsheet}) with respect to $u_s$, and then find the caustic radius $u=u_c$ where both derivatives vanish. The caustic radius $u_c$ will be a function of the starting point $u_s$; it tells us how far we should continue the ray that started at $u_s$.  
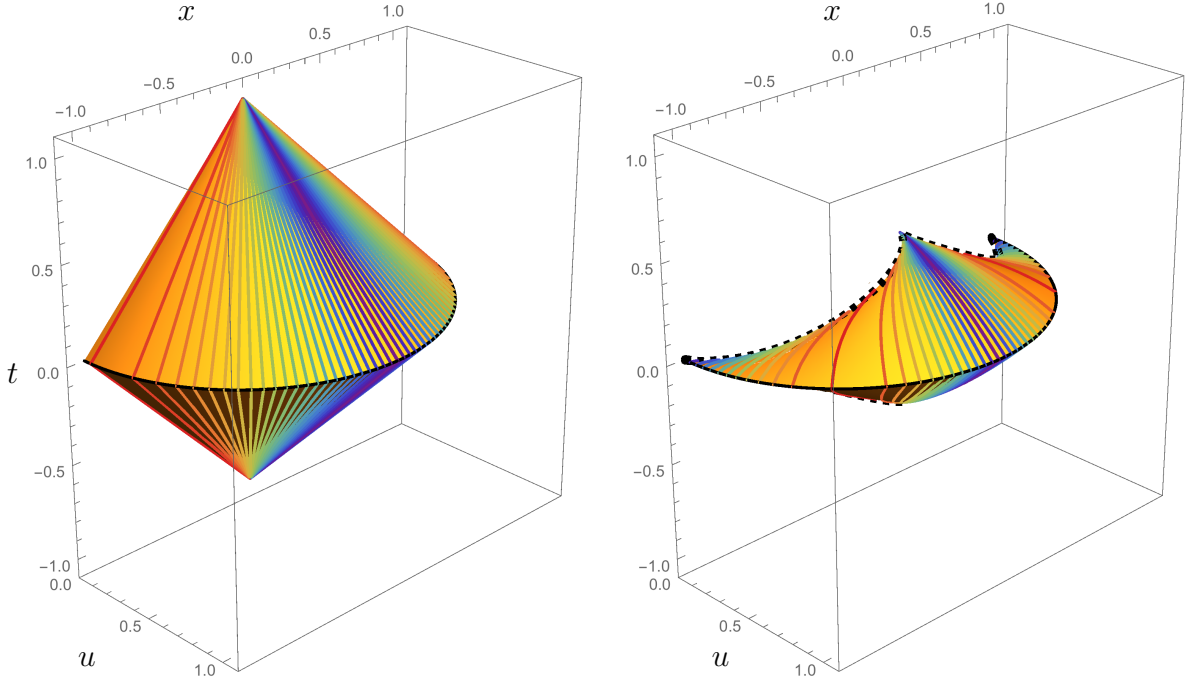
\begin{figure}[ht!]
\begin{center}
\hskip.75em
\includegraphics[width=0.45\textwidth]{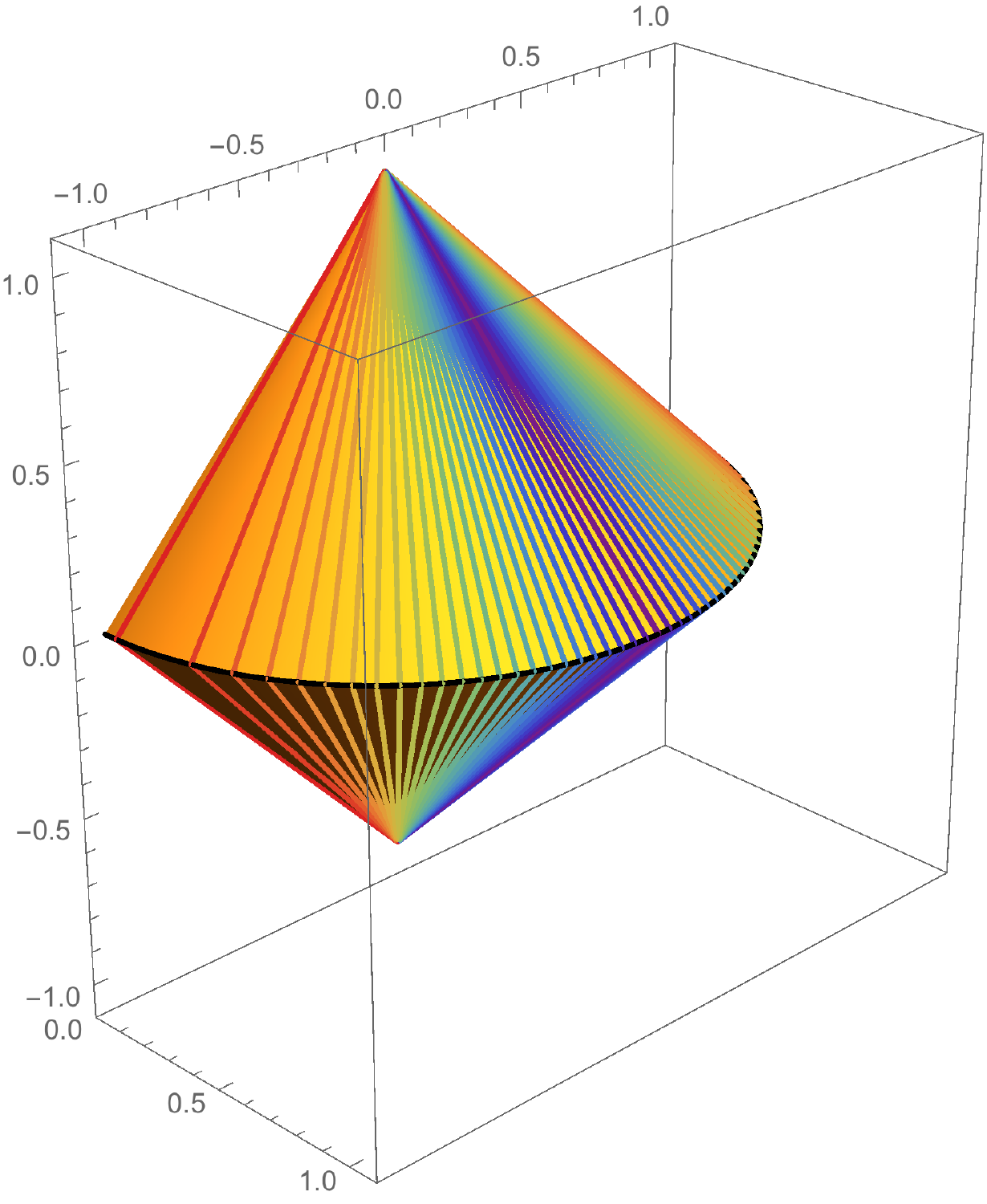}
\hskip1em
\includegraphics[width=0.45\textwidth]{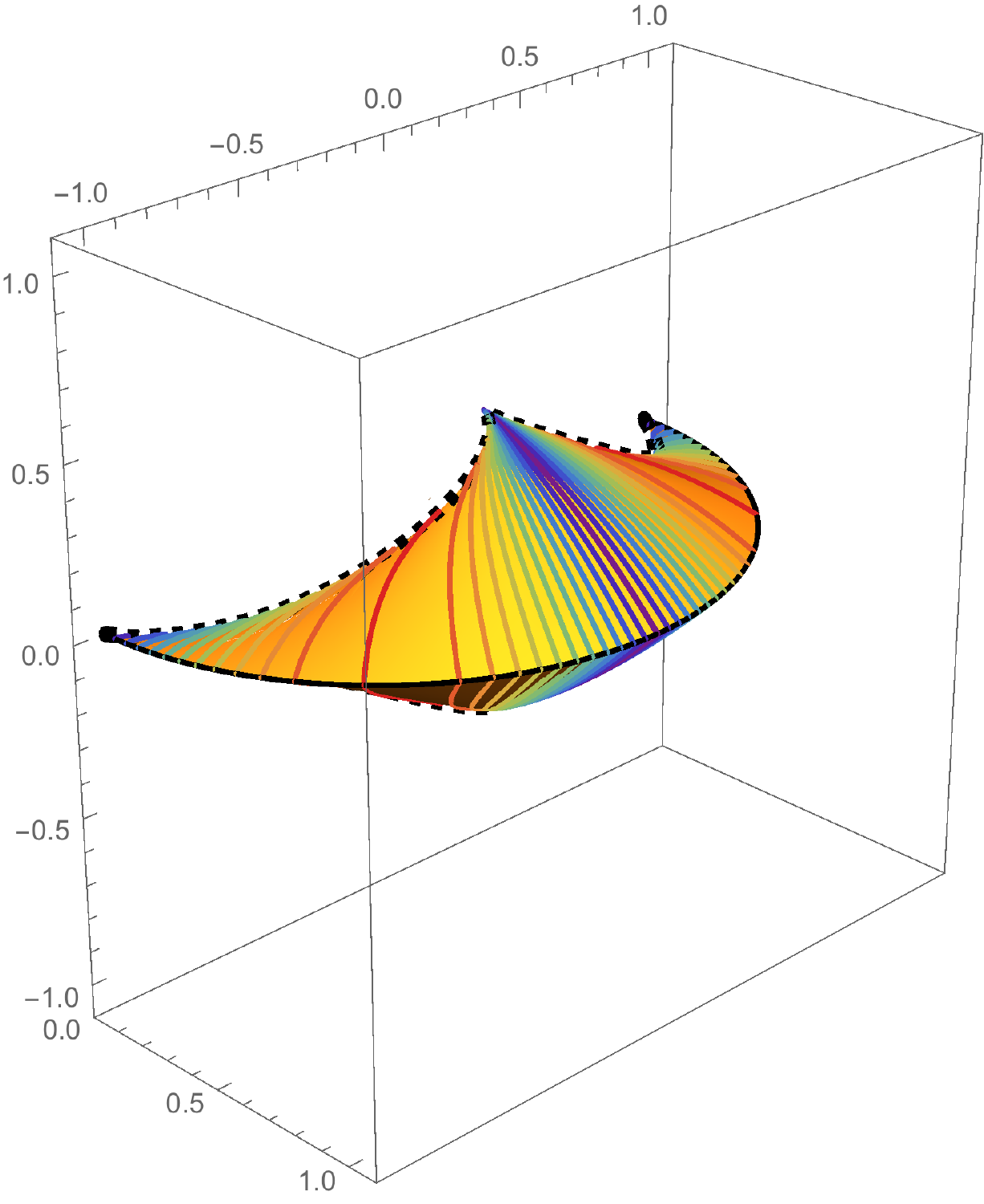}
\setlength{\unitlength}{0.1\columnwidth}
\begin{picture}(0.3,0.4)(0,0)
\put(-9.6,2.5){\makebox(0,0){$t$}}
\put(-9,0.2){\makebox(0,0){$u$}}
\put(-3.9,0.2){\makebox(0,0){$u$}}
\put(-8.2,5.4){\makebox(0,0){$x$}}
\put(-3.,5.4){\makebox(0,0){$x$}}
\end{picture}
\end{center}
\caption{On the left is the entanglement wedge for AdS, while on the right we plot the boundary-directed light-sheets emanating from the bulk extremal surface in Lifshitz with $z=2$.  Both plots have a boundary region ${\cal A}$ of radius $a=1$. Note that changing the size of the Lifshitz region does change the relative height (unlike in AdS), because time scales twice as fast as space for Lifshitz $z=2$.    The rainbow lines trace individual light rays comprising the light-sheets; their color indicates their $\ell$ value, with low $\ell$ being blue and large $\ell$ being red.  The extremal surface at $t=0$ is represented by the solid black line, whereas the dashed black line on the Lifshitz plot indicates the location of the caustics.}
\label{fig:AdSentwedge}
\end{figure}
\begin{figure}[ht!]
\begin{center}
\hskip0.75em
\includegraphics[width=0.3\textwidth]{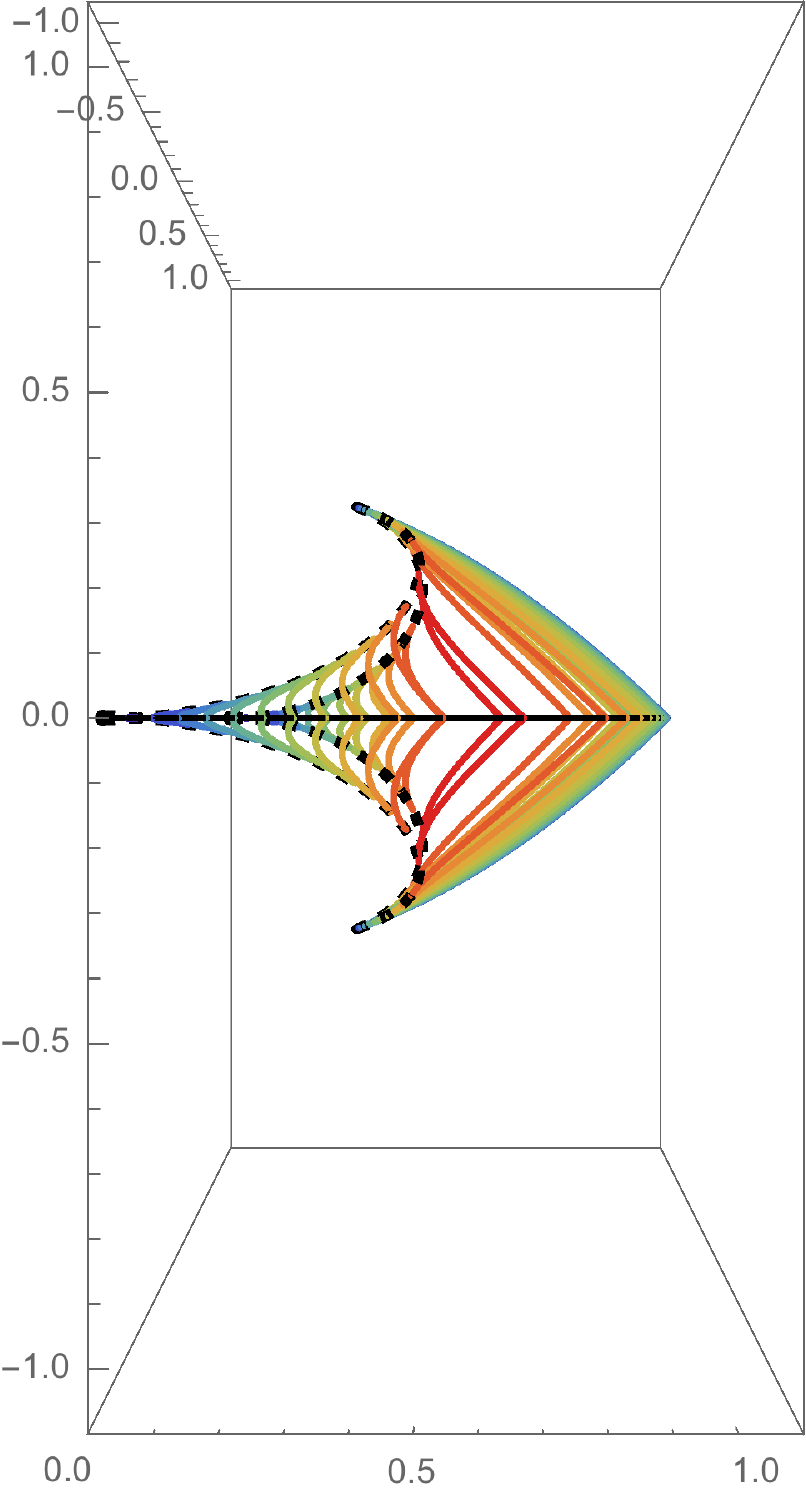}
\hskip1em
\includegraphics[width=0.3\textwidth]{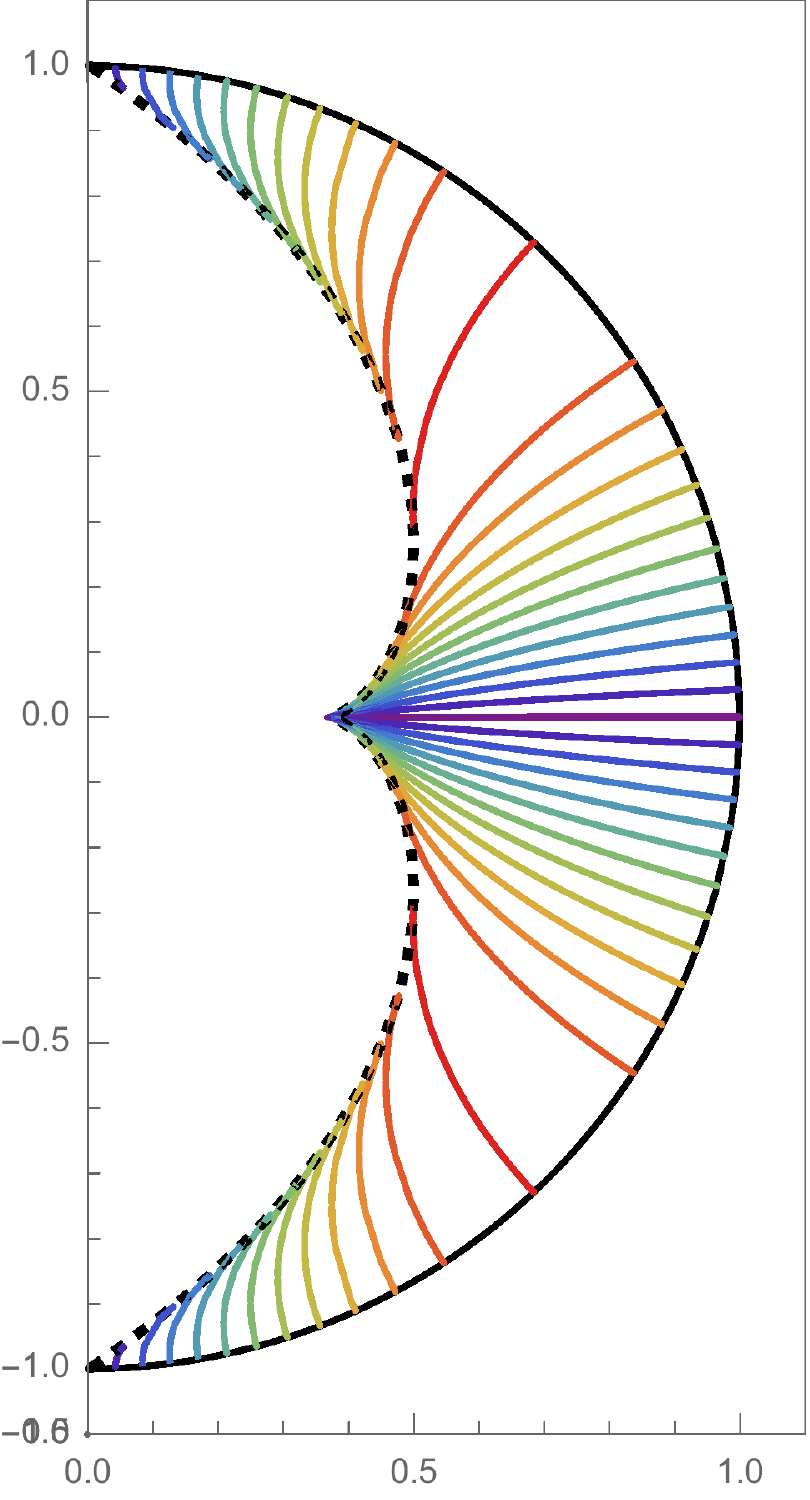}
\hskip1em
\includegraphics[width=0.3\textwidth]{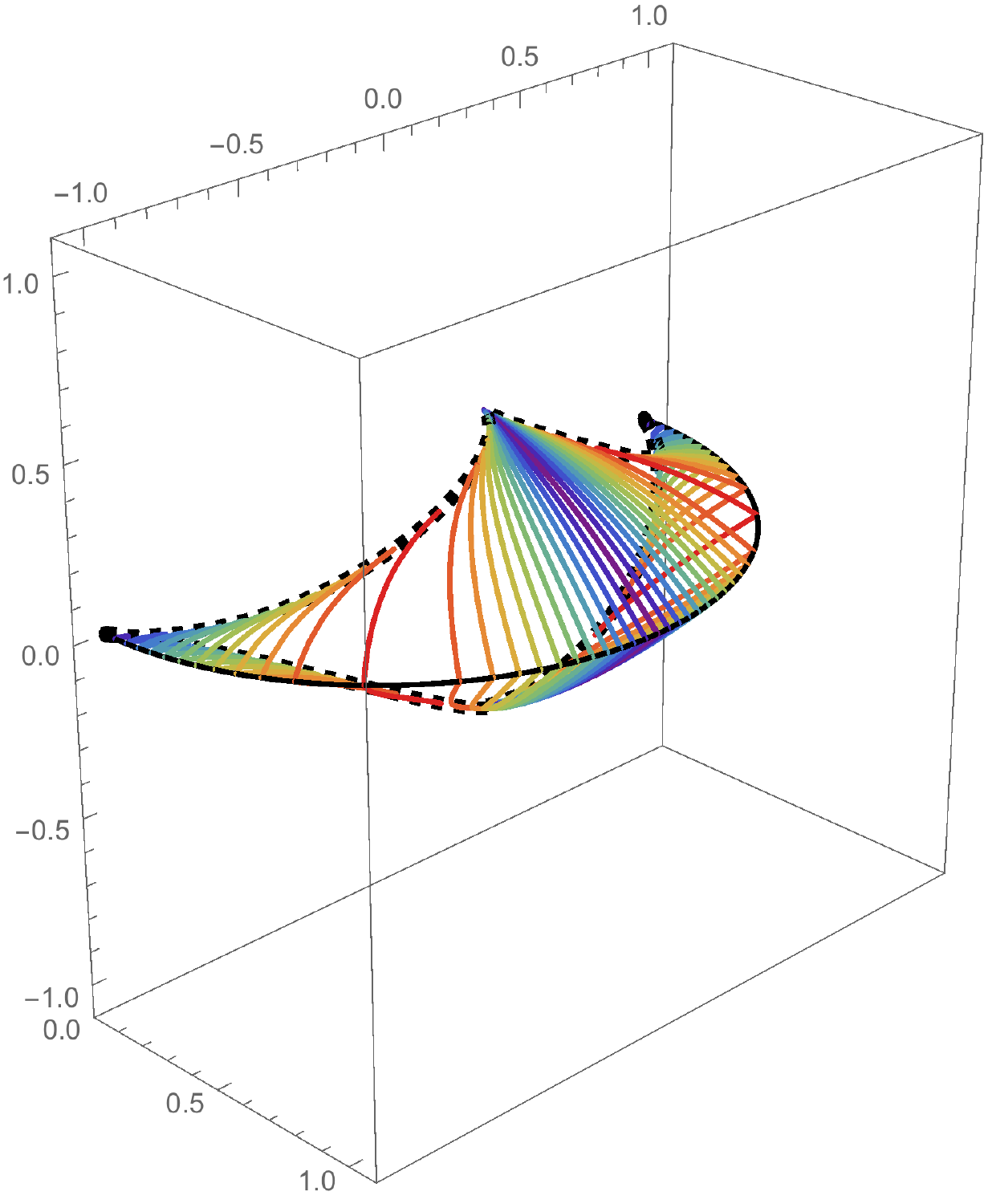}
\setlength{\unitlength}{0.1\columnwidth}
\begin{picture}(0.3,0.4)(0,0)
\put(-4.6,3.1){\makebox(0,0){$t$}}
\put(-1.2,3.1){\makebox(0,0){$x$}}
\put(.4,0.3){\makebox(0,0){$u$}}
\put(-3.1,0.3){\makebox(0,0){$u$}}
\put(-3.7,5.4){\makebox(0,0){$x$}}
\end{picture}
\end{center}
\vskip-1em
\caption{Here we have three views of the Lifshitz entanglement wedge attempt: a side view, top view and angled view. Again the colors of the light rays indicate their $\ell$, with red corresponding to large $\ell$ and purple corresponding to $\ell=0$. As is visible in the top view,
the line of caustics indicated by the dashed black line only touches the boundary at the edges of ${\cal A}$; the closest approach at $x=0$ is at $u=a/e$, $ t = (a^2/2)(1 - 1/e^2) $.
 Generic points on the line of caustics were found numerically. As we can see from \eqref{eq:ellnormalz}, the maximum $\ell$ value occurs at $u_s=a/\sqrt{2}$ here, not at the boundary. }
\label{fig:moreLifwedge}
\end{figure}

As the right hand plot in figure~\ref{fig:AdSentwedge} shows (see also figure~\ref{fig:moreLifwedge}), indeed the light-sheets built from the spacelike surface ${\cal E}_{\cal A}$ alone do not form the boundary of a bulk subregion. Although we have only shown the algebra above for the case when the boundary region ${\cal A}$ is on a constant time slice, extending to the non-constant time slice case (at a cut-off radius) does not improve the situation, nor does considering regions of different sizes.  Since the sheets do not come close to the boundary, a cut-off radius alone also cannot help.

Cutting off the light-sheets at caustics could be the source of the problem; perhaps this criterion terminates the light-sheets too early.
However, as we can see in figure~\ref{fig:pastcaustics} below, continuing to follow the null geodesics that comprise our light sheet past caustics will not allow us to form the boundary of a bulk subregion; the geodesics turn around before they hit the spacetime boundary.  
\begin{figure}[ht!]
\begin{center}
\includegraphics[width=0.5\textwidth]{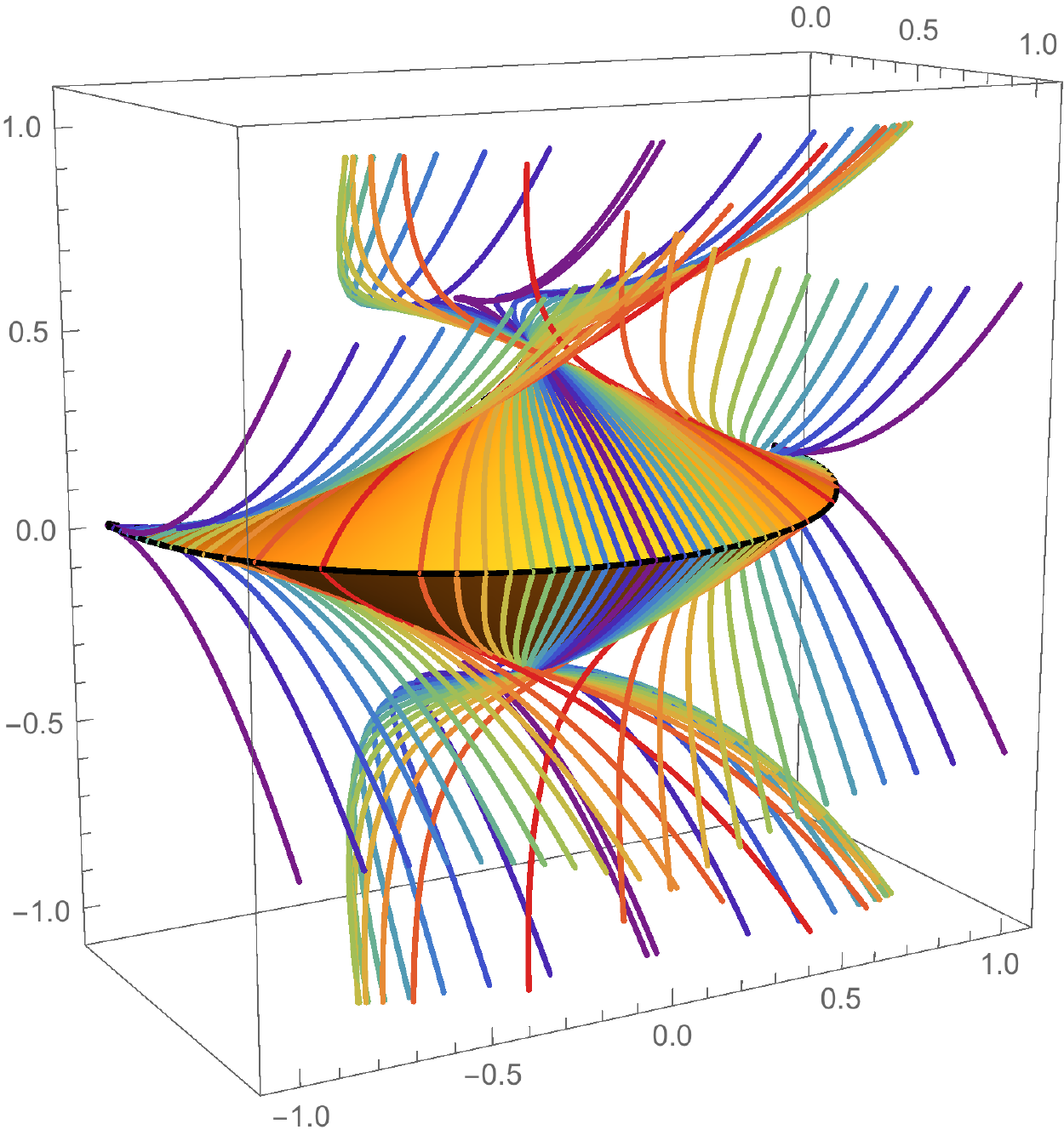}
\setlength{\unitlength}{0.1\columnwidth}
\begin{picture}(0.3,0.4)(0,0)
\put(-5.5,3.){\makebox(0,0){$t$}}
\put(-2.,0.15){\makebox(0,0){$x$}}
\put(-.6,5.5){\makebox(0,0){$u$}}
\end{picture}
\end{center}
\vskip-1em
\caption{In this angled view of the Lifshitz entanglement wedge attempt for $z=2$, continuing the light rays past caustics does not close off the wedge. Instead, all of the light rays, except the three with $\ell=0$, hit turning points and return into the bulk.}\label{fig:pastcaustics}
\end{figure}
\begin{figure}[h!]
\vskip0.6em
\begin{center}
\includegraphics[width=0.35\textwidth]{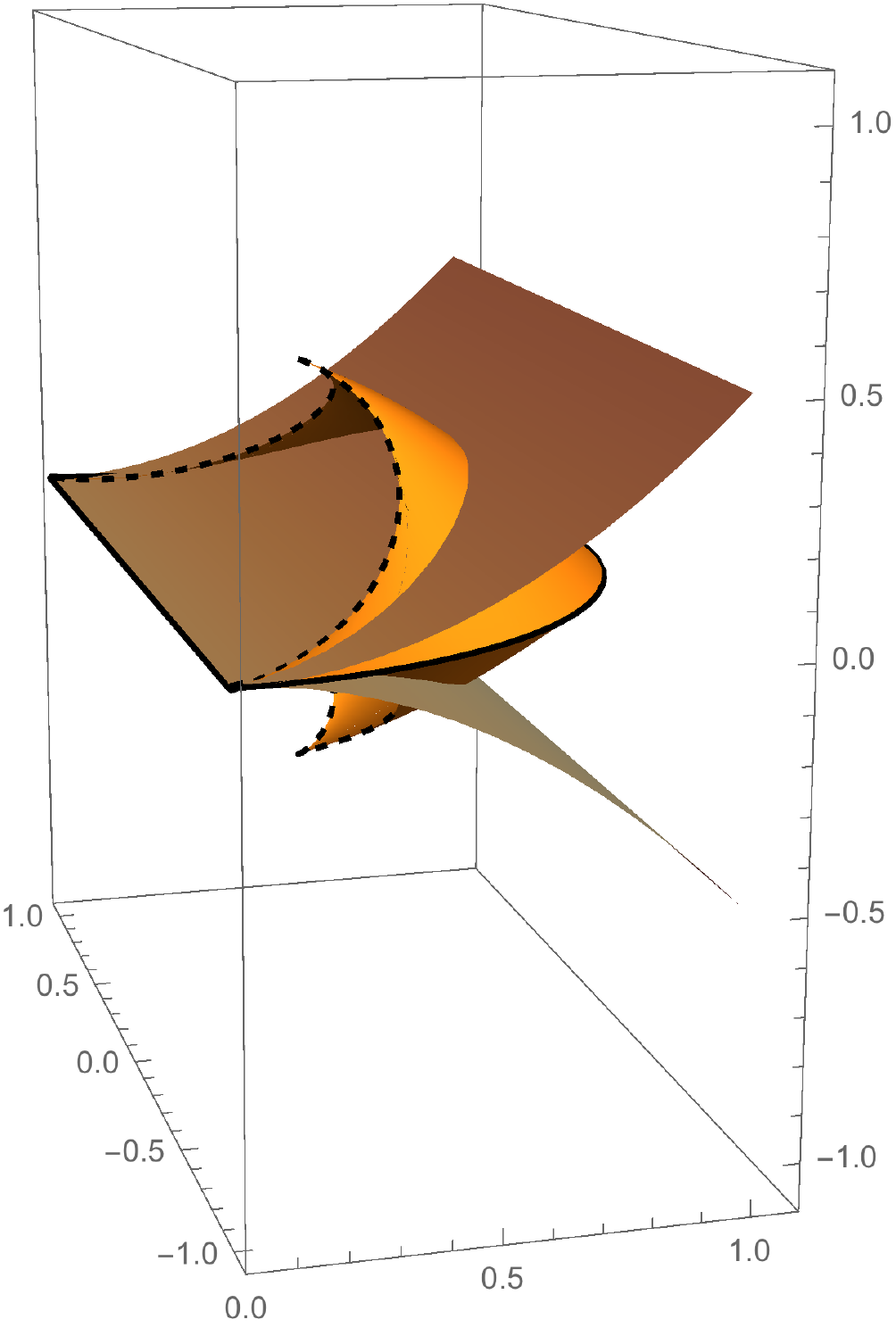}
\setlength{\unitlength}{0.1\columnwidth}
\begin{picture}(0.3,0.4)(0,0)
\put(0.1,2.7){\makebox(0,0){$t$}}
\put(-3.7,1.){\makebox(0,0){$x$}}
\put(-3.3,2.7){\makebox(0,0){${\cal A}$}}
\put(-1.6,0.0){\makebox(0,0){$u$}}
\end{picture}
\end{center}
\vskip-1em
\caption{In this angled view of the Lifshitz entanglement wedge ${\cal W}_{\cal A}$ for $z=2$, the addition of light-sheets \eqref{eq:extrasheets} (in brown) sent into the bulk from the boundary region ${\cal A}$ indeed closes the entanglement wedge.}\label{fig:actualLifentwedge}
\end{figure}

Instead, we must go back to the definition of the entanglement wedge ${\cal W}_{\cal A}$ as the causal development of the spacelike region $\Sigma$.  At the boundary $u=0$, as we discussed in the introduction, the boundary domain of dependence degenerates to become just ${\cal A}$ itself.  This occurs because the boundary theory is non-relativistic: the effective speed limit is infinite, so any point on the boundary to the future of ${\cal A}$ could be affected by events that occurred outside of ${\cal A}$ but on the same time slice.

Consequently, we also should be considering light-sheets emanating from ${\cal A}$.  For $z=2$ these light-sheets satisfy
\be\label{eq:extrasheets}
-a<x<a, \qquad t=\pm\frac{u^2}{2}.
\ee
As shown in figure~\ref{fig:actualLifentwedge}, adding in these light-sheets indeed encloses a bulk subregion.\footnote{We thank Matthew Headrick for suggesting this resolution.}

Note that adding a cut-off would complicate the shape of the light-sheets coming from ${\cal A}$, but would not remove the requirement to include both ${\cal E}_{\cal A}$- and ${\cal A}$-emanating light-sheets.  This behavior is different from AdS (and asymptotically AdS spacetimes), where the light-sheets from ${\cal E}_{\cal A}$ alone connect smoothly to the boundary causal diamond $\domd$.

The requirement to include light-sheets from ${\cal A}$ in the Lifshitz case occurs because  light-sheets leaving  ${\cal E}_{\cal A}$ in a Lifshitz spacetime do {\it not} smoothly connect to the boundary causal diamond  $\domd$ of ${\cal A}$, unlike in the AdS case. However,  once we include the light-sheet from $\cal A$, the full entanglement wedge boundary does indeed intersect the boundary of $\domd$, which is just the two endpoints of $\cal A$ in this case. We note that this smooth matching of the light-sheet from ${\cal E}_{\cal A}$ {\it alone} in the AdS case was used by HRT \cite{Hubeny:2007xt} as partial justification for their   covariant holographic entanglement entropy proposal.

\section{Discussion}\label{sec:Discussion}%

As we have shown, spacetime reconstruction techniques valid for asymptotically AdS spacetimes require re-evaluation for asymptotically Lifshitz backgrounds.  We find that the changes in the bulk causal structure for $z>1$ as well as the non-relativistic boundary behavior both serve to alter the reconstructibility of bulk spacelike curves, as well as the identification of the bulk reconstructible region via either the causal wedge or entanglement wedge.  Constructions that work simply in AdS may encounter complications in Lifshitz systems.

The differential entropy approach to building spacelike bulk curves succeeds for constant radius curves, when the bulk curve has tangents satisfying \eqref{eq:constructible}. Indeed, even the cut-off necessary to define the differential entropy drops out of the final result.  If the curve takes on a larger value for $T'(\lambda)$ anywhere along its path at $u=u_\star$, then we cannot reconstruct it from a series of boundary-anchored curves, for the simple reason that the tangent extremal curve is not boundary-anchored.  Crucially, it is possible to a pick a curve that is entirely spacelike, but cannot be reconstructed, for any given radius.

For the causal wedge we also found partial success: we are able to define a causal wedge based on a regulated domain of dependence.  However both the height and the bulk depth of the wedge shrink to zero as we remove the cut-off.  If the causal wedge correctly identifies the bulk region reconstructible from the information within the starting boundary region, then this degeneration may indicate that only with a strict cut-off can we hope to reconstruct any of the bulk.

For the entanglement wedge, the problem is different.  The light-sheets from the holographic entanglement extremal surface do not reach the boundary and thus do not enclose a bulk region.  Continuing the light-sheets past caustics does not fix the problem, since light rays with transverse momentum in Lifshitz spacetime turn around before reaching the boundary.  For an interval of width $2a$ in  Lif$_3$ with $z=2$, the light rays have momentum as high as $\ell=a/2$, which corresponds to a geodesic that turns around at $u=a/2$,  so a small cut-off alone cannot close the wedge. Instead, we must additionally include light-sheets emanating from the boundary region ${\cal A}$: these light-sheets combined with those from the extremal surface ${\cal E}_{\cal A}$ do form a closed subregion of the bulk.

 This change in the formation of the entanglement wedge does not alter the identification of intersections of entanglement wedges with the spacelike curve reconstructed via the differential entropy, as mentioned at the end of section~\ref{sec:differentialentropy}.  Although the light-sheets emanating from a single extremal surface do not enclose a bulk subregion, a pair of neighboring light-sheets still may. 
 
 However, the change in the formation of the entanglement wedge does indicate a concern regarding the HRT proposal in a Lifshitz spacetime. One justification of this proposal begins with a series of past and future light-sheets that asymptotically approach the past and future of the boundary causal diamond; where each pair of light sheets intersects, HRT define a candidate surface.  Among these candidate surfaces, their proposal picks out the surface that extremizes its area, as justified by the Bousso entropy bound.  In the Lifshitz case, it is still possible to construct many light-sheets that do asymptote to the boundary causal diamond in asymptotically Lifshitz spacetime; however the light-sheets emanating from the spacelike geodesic corresponding to the Ryu-Takayangi or HRT proposal are not in this class.   If we insist on considering light-sheets that asymptote to the boundary causal domain, we would then need another prescription to decide between the many possible light-sheets.  
 
We have limited our discussion here to vacuum Lifshitz spacetimes in three dimensions. For the differential entropy, this means both that we study only spacelike geodesics rather than higher-dimensional extremal surfaces, and also that there is only one spacelike geodesic homologous to a given boundary region.  From our experience with non-vacuum AdS spacetimes, we expect that considering instead just asymptotically Lifshitz spacetimes may require choosing between two (or more) extremal surfaces. 

For the causal wedge, our restriction to three dimensions means that the boundary domain of dependence is fairly trivial: domains of dependence in two dimensions are necessarily composed entirely of causal diamonds.  In higher dimensions the boundary domain may be more complicated, but we do not believe this will alter our result; we will still require a cut-off surface in order to define a non-degenerate wedge, but with a cut-off the construction should be successful.

We are left with many unresolved questions.  First, naive application of the Ryu-Takayanagi proposal to Lifshitz spacetimes may not be justified; there are few field theory calculations to compare to.  The lack of dependence on the dynamical exponent $z$ in the vacuum could be physical but may also be indicative of a required modification of this proposal.  It would be interesting to see if the  justification of this proposal given in \cite{Lewkowycz:2013nqa} can be extended to asymptotically Lifshitz spacetimes.

In order to study dependence on $z$ more thoroughly we could consider boundary regions of non-constant time, defined on a cut-off surface.  However it is  unclear to us what meaning an entanglement entropy on such a  region would have in the boundary field theory, much less that said quantity should be associated with the area of a homologous extremal surface.  In fact, if we wish to consider boundary regions at something other than constant $t$, we should instead consider the field theory symmetry algebra directly, rather than just inheriting the relativistic bulk symmetry at a cut-off surface (see for example \cite{Bagchi:2014iea, Hosseini:2015uba, Castro:2015csg}). Perhaps a Galilean boosted region might be more interesting to consider.

We should note that the lack of physical interpretation in the boundary field theory for the entanglement entropy of a boosted region means that even curves that can be successfully reconstructed via a differential entropy approach have not really been built from boundary information; or rather, we do not understand the physical boundary information required to compute the length of such curves if they are not on a constant time slice.  This problem, coupled with the lack of field-theoretic calculations for Lifshitz field theories, also precludes using the variational approach of \cite{Lashkari:2013koa, Faulkner:2013ica} to rebuild Lifshitz spacetime. Given the differences between Lifshitz and AdS that we have exhibited here, we expect that other entanglement-reconstruction ides, including universal properties of entanglement reconstruction \cite{Haehl:2015rza} and entanglement holography \cite{deBoer:2015kda}, will also be altered when studied for non-relativistic duals.

Even outside of the question of asymptotically Lifshitz spacetimes, some of our results also affect asymptotically AdS spacetimes, provided they have a Lifshitz-like interior.  For example, reconstruction of bulk curves deep in a Lifshitz-like region may fail due to a lack of boundary-anchored extremal surfaces.  Such problems may additionally generalize to other spacetimes with light rays that do not reach the boundary, such as black holes in AdS. This possible generalization, which remains to be explored fully, is evidenced by the connection between our work and that of \cite{Engelhardt:2015dta}.

It would be interesting to explore the relationship between boundary and bulk causality in an asymptotically Lifshitz spacetime.  In asymptotically AdS spacetimes, the gravitational time-delay theorem of Gao and Wald~\cite{Gao:2000ga} implies that bulk causality agrees  with boundary causality for local CFT observables.  (In brief: a signal connecting two boundary points  cannot propagate more quickly  through the bulk than along the boundary.) It was  proven in~\cite{Headrick:2014cta} that the HRT prescription for  entanglement entropy (a non-local quantity) also satisfies this basic consistency requirement.  Both results assume that the bulk obeys the null energy condition.  Whilst many theories admitting asymptotically Lifshitz solutions satisfy this condition, the degeneracy of the Lifshitz boundary and its presumed non-relativistic dual makes this issue very different to the AdS case.  One possible in-road to this subject is to study the Lieb-Robinson bound~\cite{Lieb:1972wy} on the speed of entanglement propagation for non-relativistic systems holographically.

There is a stronger possible interpretation of the difficulties with bulk reconstruction in Lifshitz spacetimes: perhaps they arise not from problems with holographic entanglement entropy prescriptions, but instead because we are attempting to construct the wrong spacetime. In order to reconstruct a bulk dual geometry to non-relativistic theories, perhaps we should begin by studying the boundary symmetries, viewing the field theory in a Newton-Cartan approach as in \cite{Hartong:2014pma}.  Then we may hope to construct a new spacetime dual that we are able to more fully reconstruct,  as done for warped CFTs in \cite{Hofman:2014loa,Castro:2015csg}.

\section*{Acknowledgements}%

We are delighted to thank Gino Knodel, Per Kraus, James Liu, Niels Obers,  Simon Ross, Tadashi Takayanagi, Larus Thorlacius,  William Witczak-Krempa and especially Matthew Headrick for useful discussions.  SAG is supported  by National Science Foundation grant PHY-13-13986. CK is supported by the European Union's Horizon 2020 research and innovation programme under the Marie Sk\l{}odowska-Curie grant agreement No 656900.


\appendix

\section{A useful null congruence}
\label{appendix:screens}
Recently, \cite{Engelhardt:2015dta} proved that any spacetime with a  holographic screen contains curves that are not constructible via the hole-ography or differential entropy approaches.  In this section, we demonstrate a class of null congruences possessing almost-holographic screens within the vacuum Lifshitz spacetime.  Consequently, despite the success of the approach for the specific curves in (\ref{eq:gammaB}), (\ref{eq:periodicBCgammaB}), we are not surprised that vacuum Lifshitz spacetime must contain curves for which the differential entropy reconstruction approach fails.  Additionally, we use Lemma 2 from \cite{Engelhardt:2015dta} to show that spacelike geodesics with radii above (\ref{eq:umaxgenz}) never reach a smaller $u$ and thus cannot touch the boundary or participate in the hole-ography approach.

To define a holographic screen, \cite{Engelhardt:2015dta} begin with a null foliation of (a region of) the spacetime, defined via the foliation's null generator $k^\mu$. We will assume $k^\mu$ is future directed, and use $\theta_k$ to refer to the null expansion. 
Along each codimension-one leaf of the foliation, we build a sequence of co-dimension two surfaces $\sigma$ by choosing another null vector $l^\mu$ tangent to the leaf.  These `leaflets'%
\footnote{We follow the terminology and notation in \cite{Engelhardt:2015dta}.} %
 $\sigma$ have two null normal directions, $k^\mu$ and $l^\mu$. Let $\sigma_N$ be the location on an individual leaf $N$ of this foliation where the null congruence's expansion vanishes; this can happen at most once per foliation as long as the null energy condition holds.

A holographic screen is the union of these $\sigma_N$ across the entire foliation, assuming that the expansion $\theta_l$ is non-zero.  It is a future holographic screen if $\theta_l <0$, and a past holographic screen if $\theta_l >0$.  In our case below, however, we will find $\theta_k$ and $\theta_l$ vanish at the same location.

Even though the Theorems in \cite{Engelhardt:2015dta} are not applicable because both $\theta_k$ and $\theta_l$ vanish at the same location, we can still use Lemma 2 (conveniently applicable in 2+1 dimensions!) to show that geodesics with turning radii between (\ref{eq:umaxgenz}) and the spacelike limit $(P/E)^{1/(1-z)}$ cannot reach the boundary. Lemma 2 shows that in a region of a null congruence with $\theta_k <0$, for a given leaf $N$ of the associated foliation, any spacelike geodesic $X$ that is tangent to the leaf $N$ at a point $p$ has some neighborhood ${\cal O}_p$ such that $X \cap {\cal O}_p$ is nowhere to the past of~$N$.

Let us now be explicit for the case of Lifshitz.  We consider the foliation built from the null generator
\begin{equation}\label{eq:keqn}
k^\mu = \{u^{2z}, u \sqrt{u^{2z}-u^2 \rho^2}, \rho u^2\},
\end{equation}
with $\rho$ a positive constant. Individual leaves of this foliation solve
\begin{equation}
t=\rho x + \int du k_u + c_N,
\end{equation}
where the constant $c_N$ specifies the leaf in the foliation, and we assume $\rho>0$ for simplicity.  This foliation is spacetime filling for $u>\rho^{1/(z-1)}$.

It is easiest to compute $\theta_k$ by choosing another non-aligned null vector; the choice of null vector is immaterial but we will choose one suited to the $\sigma$ above. Specifically, we choose
\begin{equation}
l^\mu = \{u^{2z}, -u \sqrt{u^{2z}-u^2 \rho^2}, \rho u^2\}.
\end{equation}
Note that we do not have $k\cdot l = -1$, but this could be acheived by a trivial rescaling of $l^\mu$ if needed.  We will not need it here.

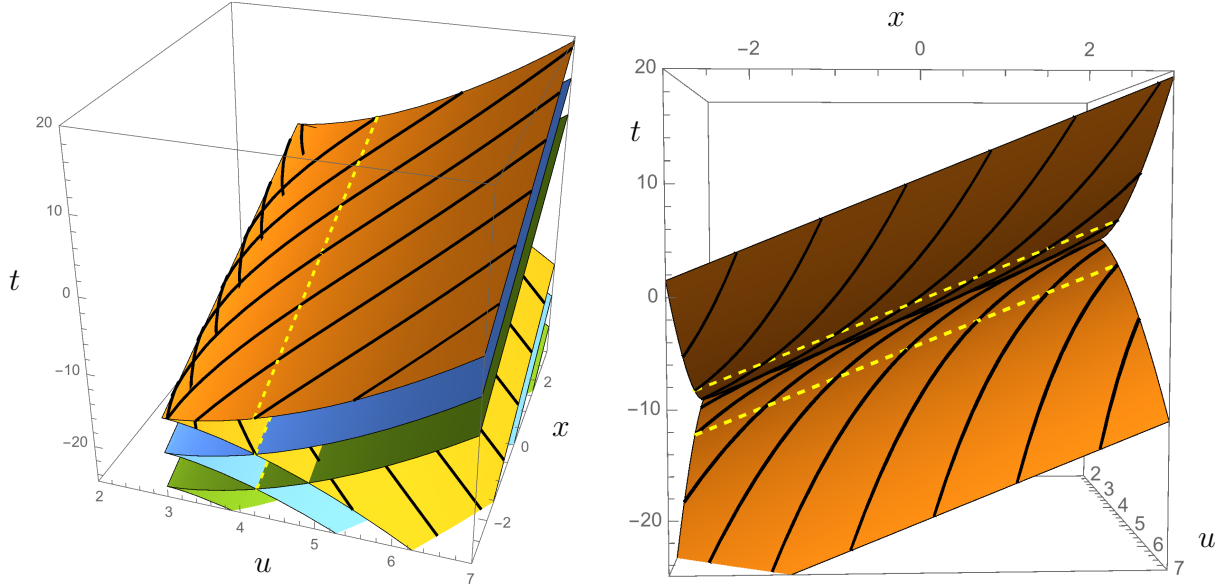
\begin{figure}[h!]
\begin{center}
\hskip0.75em
\includegraphics[width=0.45\textwidth]{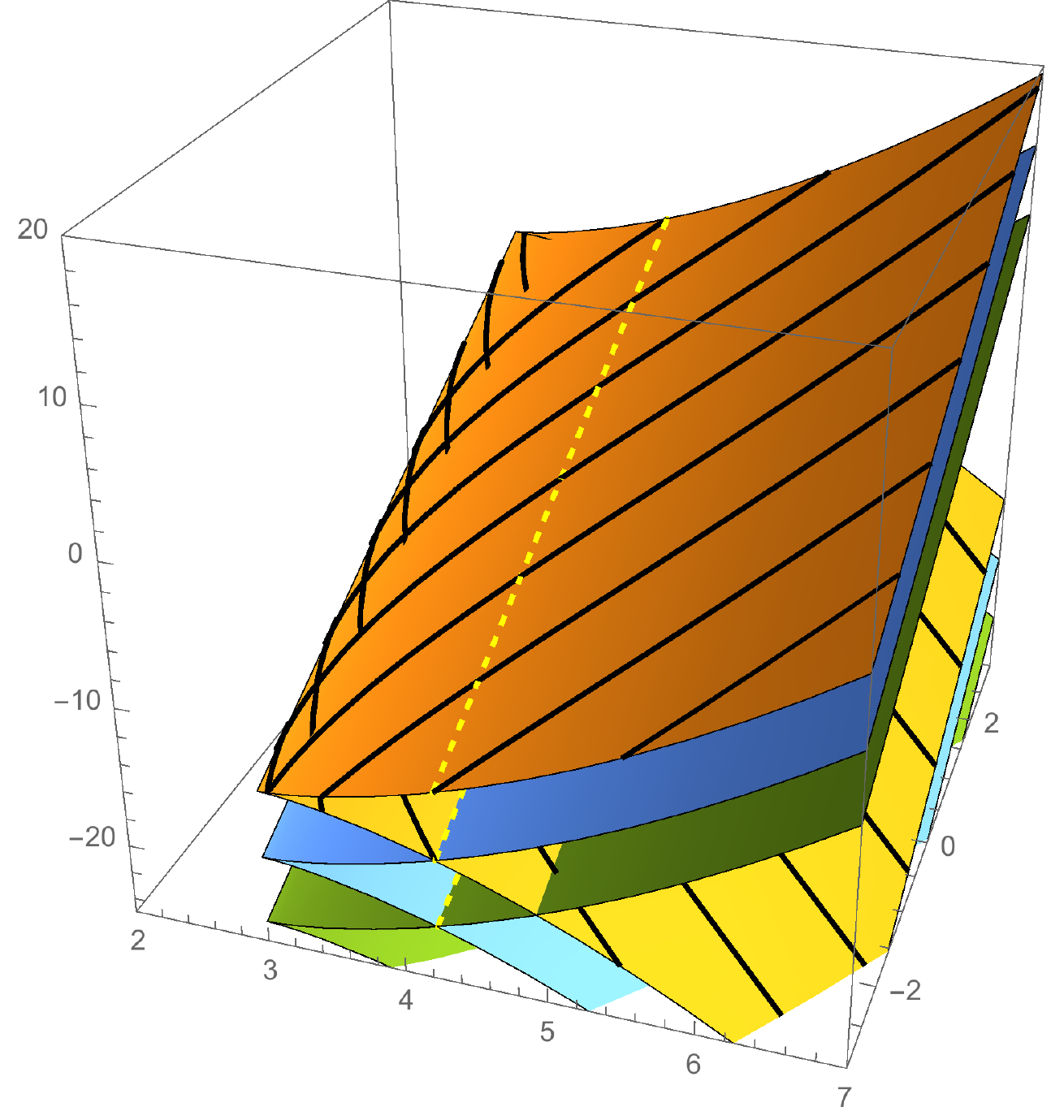}
\hskip1em
\includegraphics[width=0.45\textwidth]{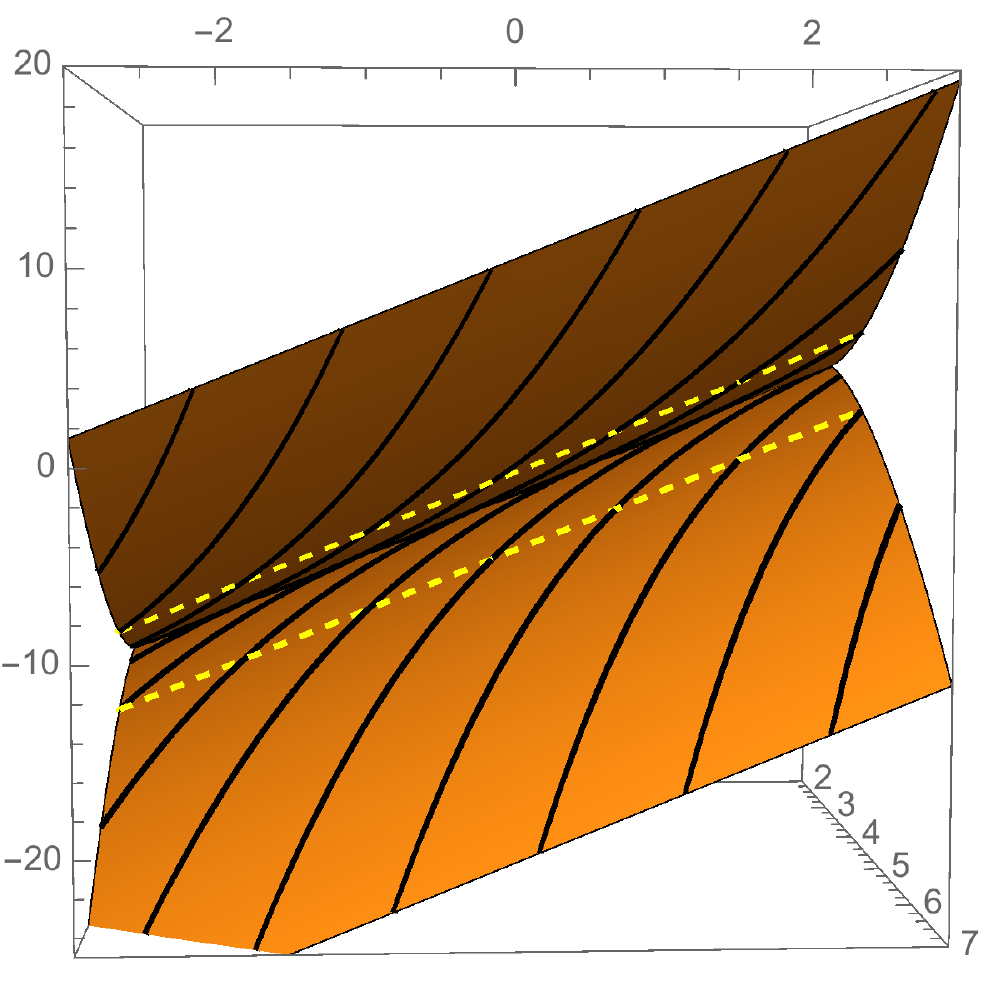}
\setlength{\unitlength}{0.1\columnwidth}
\begin{picture}(0.3,0.4)(0,0)
\put(-9.6,2.5){\makebox(0,0){$t$}}
\put(-5.2,1.3){\makebox(0,0){$x$}}
\put(-4.6,3.7){\makebox(0,0){$t$}}
\put(0.0,0.4){\makebox(0,0){$u$}}
\put(-2.5,4.6){\makebox(0,0){$x$}}
\put(-7.6,0.2){\makebox(0,0){$u$}}
\end{picture}
\end{center}
\caption{Representations of the congruence $k^\mu$, for $\rho=3$.  Different colors denote different values of $c_N$, where the upper leaves of each color correspond to the $k^\mu$ congruence and the lower ones correspond to $\ell^\mu$.  The light rays comprising the top pair of leaves have been drawn in black; we can see they have a turning point where the leaves join.  Both congruences have expansion $\theta=0$ at the dotted lines where $u=u_N$.  Spacelike geodesics tangent to the upper leaf for $u<u_N$ must be to the past of, or below, the leaf; that is, they head away from the boundary.  Conversely, spacelike geodesics tangent to the upper leaf for $u>u_N$ must be to the future of, or above, the leaf, so they head in towards the boundary.}\label{fig:congruence}
\end{figure}

Instead, we now compute $\theta_k$ by first defining the reduced induced metric $h^{\mu\nu}$:
\begin{equation}
h^{\mu\nu} \equiv g^{\mu\nu} +k^\mu l^\nu +l^\mu k^\nu,
\end{equation}
and then computing 
\begin{equation}
h^{\mu\nu} \nabla_\mu k_\nu \equiv \theta_k = \frac{\rho^2 u^2 z - u^{2z}}{\sqrt{u^{2z} - \rho^2 u^2}}.
\end{equation}
Similarly for the expansion of $l^\mu$ we find
\begin{equation}
h^{\mu\nu} \nabla_\mu l_\nu \equiv \theta_l = - \frac{\rho^2 u^2 z - u^{2z}}{\sqrt{u^{2z} - \rho^2 u^2}}.
\end{equation}
In both cases, the $\sigma_N$, where $\theta=0$, occur at constant radius $u_N$ given by
\begin{equation}\label{eq:uNeqn}
u_N = \left(\rho^2 z\right)^{\frac{1}{2(z-1)}}.
\end{equation}
Note for $k$, the expansion $\theta_k$ becomes negative for $u>u_N$.  This makes sense as $k$ is future-directed and headed towards larger radius at larger $t$, so indeed $\theta_k$ can only decrease (or stay constant) as $u$ gets larger.  It is also positive for $u<u_N$.
Thus by Lemma 2 of \cite{Engelhardt:2015dta}, any spacelike geodesic tangent to this foliation at $u>u_N$ must be to the future of $N$ (for a small neighborhood).  See figure~\ref{fig:congruence} for illustration.

Now, let us consider how this result affects the spacelike geodesics considered in section~\ref{reconstruction}.  For simplicity we will explicitly consider only geodesics that are tangent to the foliation at their radial turning point.  This means their tangent vector at the radial turning point should solve $T^\mu k_\mu =0$, with $T^\mu$ given as in (\ref{eq:Edef}) and (\ref{eq:Pdef}), so we require
\begin{equation}\label{eq:findrho}
\rho = \frac{E u_\star^{2(z-1)}}{P} = \frac{T'(\lambda)}{\xi}
\end{equation}
where here $u_\star$ is the radial turning point. The second equality comes from \eqref{eq:tdotoverxdot}, which relates the spacelike curve we are trying to rebuild to its tangent geodesic.

Consequently, in order to see if a particular spacelike curve is reconstructible, we should examine the congruence given by \eqref{eq:keqn} with $\rho$ satisfying \eqref{eq:findrho}.  If the radial location of the curve we are trying to reconstruct, $u_\star$, is smaller than $u_N$ as in \eqref{eq:uNeqn}, then the extremal curve through it will not reach the boundary.  However, if $u_\star>u_N$, then the curve can be constructed.  Rewriting using \eqref{eq:findrho}, we find the curve is constructible as long as
\begin{equation}\label{eq:constructible}
u_\star^{2(z-1)} > \left(\frac{T'(\lambda)}{\xi}\right)^2 z.
\end{equation}
%



\begin{thebibliography}{10}

\bibitem{Headrick:2014eia}
M.~Headrick, R.~C. Myers, and J.~Wien, ``{Holographic Holes and Differential
  Entropy},'' \href{http://dx.doi.org/10.1007/JHEP10(2014)149}{{\em JHEP}
  {\bfseries 1410} (2014) 149},
\href{http://arxiv.org/abs/1408.4770}{{\ttfamily arXiv:1408.4770 [hep-th]}}.

\bibitem{Balasubramanian:1999ri}
V.~Balasubramanian, S.~B. Giddings, and A.~E. Lawrence, ``{What do CFTs tell us
  about Anti-de Sitter space-times?},''
  \href{http://dx.doi.org/10.1088/1126-6708/1999/03/001}{{\em JHEP} {\bfseries
  03} (1999) 001},
\href{http://arxiv.org/abs/hep-th/9902052}{{\ttfamily arXiv:hep-th/9902052
  [hep-th]}}.

\bibitem{Freivogel:2013zta}
B.~Freivogel and B.~Mosk, ``{Properties of Causal Holographic Information},''
  \href{http://dx.doi.org/10.1007/JHEP09(2013)100}{{\em JHEP} {\bfseries 1309}
  (2013) 100},
\href{http://arxiv.org/abs/1304.7229}{{\ttfamily arXiv:1304.7229 [hep-th]}}.

\bibitem{Keeler:2013msa}
C.~Keeler, G.~Knodel, and J.~T. Liu, ``{What do non-relativistic CFTs tell us
  about Lifshitz spacetimes?},''
  \href{http://dx.doi.org/10.1007/JHEP01(2014)062}{{\em JHEP} {\bfseries 01}
  (2014) 062},
\href{http://arxiv.org/abs/1308.5689}{{\ttfamily arXiv:1308.5689 [hep-th]}}.

\bibitem{Keeler:2014lia}
C.~Keeler, G.~Knodel, and J.~T. Liu, ``{Hidden horizons in non-relativistic
  AdS/CFT},'' \href{http://dx.doi.org/10.1007/JHEP08(2014)024}{{\em JHEP}
  {\bfseries 1408} (2014) 024},
\href{http://arxiv.org/abs/1404.4877}{{\ttfamily arXiv:1404.4877 [hep-th]}}.

\bibitem{Ryu:2006ef}
S.~Ryu and T.~Takayanagi, ``{Aspects of Holographic Entanglement Entropy},''
  \href{http://dx.doi.org/10.1088/1126-6708/2006/08/045}{{\em JHEP} {\bfseries
  08} (2006) 045},
\href{http://arxiv.org/abs/hep-th/0605073}{{\ttfamily arXiv:hep-th/0605073
  [hep-th]}}.

\bibitem{Ryu:2006bv}
S.~Ryu and T.~Takayanagi, ``{Holographic derivation of entanglement entropy
  from AdS/CFT},'' \href{http://dx.doi.org/10.1103/PhysRevLett.96.181602}{{\em
  Phys. Rev. Lett.} {\bfseries 96} (2006) 181602},
\href{http://arxiv.org/abs/hep-th/0603001}{{\ttfamily arXiv:hep-th/0603001
  [hep-th]}}.

\bibitem{Hubeny:2007xt}
V.~E. Hubeny, M.~Rangamani, and T.~Takayanagi, ``{A Covariant holographic
  entanglement entropy proposal},''
  \href{http://dx.doi.org/10.1088/1126-6708/2007/07/062}{{\em JHEP} {\bfseries
  07} (2007) 062},
\href{http://arxiv.org/abs/0705.0016}{{\ttfamily arXiv:0705.0016 [hep-th]}}.

\bibitem{Lashkari:2013koa}
N.~Lashkari, M.~B. McDermott, and M.~Van~Raamsdonk, ``{Gravitational dynamics
  from entanglement `thermodynamics'},''
  \href{http://dx.doi.org/10.1007/JHEP04(2014)195}{{\em JHEP} {\bfseries 04}
  (2014) 195},
\href{http://arxiv.org/abs/1308.3716}{{\ttfamily arXiv:1308.3716 [hep-th]}}.

\bibitem{Faulkner:2013ica}
T.~Faulkner, M.~Guica, T.~Hartman, R.~C. Myers, and M.~Van~Raamsdonk,
  ``{Gravitation from Entanglement in Holographic CFTs},''
  \href{http://dx.doi.org/10.1007/JHEP03(2014)051}{{\em JHEP} {\bfseries 03}
  (2014) 051},
\href{http://arxiv.org/abs/1312.7856}{{\ttfamily arXiv:1312.7856 [hep-th]}}.

\bibitem{Hammersley:2006cp}
J.~Hammersley, ``{Extracting the bulk metric from boundary information in
  asymptotically AdS spacetimes},''
  \href{http://dx.doi.org/10.1088/1126-6708/2006/12/047}{{\em JHEP} {\bfseries
  12} (2006) 047},
\href{http://arxiv.org/abs/hep-th/0609202}{{\ttfamily arXiv:hep-th/0609202
  [hep-th]}}.

\bibitem{Hammersley:2007ab}
J.~Hammersley, ``{Numerical metric extraction in AdS/CFT},''
  \href{http://dx.doi.org/10.1007/s10714-007-0564-6}{{\em Gen. Rel. Grav.}
  {\bfseries 40} (2008) 1619--1652},
\href{http://arxiv.org/abs/0705.0159}{{\ttfamily arXiv:0705.0159 [hep-th]}}.

\bibitem{Bilson:2008ab}
S.~Bilson, ``{Extracting spacetimes using the AdS/CFT conjecture},''
  \href{http://dx.doi.org/10.1088/1126-6708/2008/08/073}{{\em JHEP} {\bfseries
  08} (2008) 073},
\href{http://arxiv.org/abs/0807.3695}{{\ttfamily arXiv:0807.3695 [hep-th]}}.

\bibitem{Bilson:2010ff}
S.~Bilson, ``{Extracting Spacetimes using the AdS/CFT Conjecture: Part II},''
  \href{http://dx.doi.org/10.1007/JHEP02(2011)050}{{\em JHEP} {\bfseries 02}
  (2011) 050},
\href{http://arxiv.org/abs/1012.1812}{{\ttfamily arXiv:1012.1812 [hep-th]}}.

\bibitem{Balasubramanian:2013lsa}
V.~Balasubramanian, B.~D. Chowdhury, B.~Czech, J.~de~Boer, and M.~P. Heller,
  ``{Bulk curves from boundary data in holography},''
  \href{http://dx.doi.org/10.1103/PhysRevD.89.086004}{{\em Phys. Rev.}
  {\bfseries D89} no.~8, (2014) 086004},
\href{http://arxiv.org/abs/1310.4204}{{\ttfamily arXiv:1310.4204 [hep-th]}}.

\bibitem{Myers:2014jia}
R.~C. Myers, J.~Rao, and S.~Sugishita, ``{Holographic Holes in Higher
  Dimensions},'' \href{http://dx.doi.org/10.1007/JHEP06(2014)044}{{\em JHEP}
  {\bfseries 06} (2014) 044},
\href{http://arxiv.org/abs/1403.3416}{{\ttfamily arXiv:1403.3416 [hep-th]}}.

\bibitem{Czech:2012bh}
B.~Czech, J.~L. Karczmarek, F.~Nogueira, and M.~Van~Raamsdonk, ``{The Gravity
  Dual of a Density Matrix},''
  \href{http://dx.doi.org/10.1088/0264-9381/29/15/155009}{{\em Class. Quant.
  Grav.} {\bfseries 29} (2012) 155009},
\href{http://arxiv.org/abs/1204.1330}{{\ttfamily arXiv:1204.1330 [hep-th]}}.

\bibitem{Hubeny:2012wa}
V.~E. Hubeny and M.~Rangamani, ``{Causal Holographic Information},''
  \href{http://dx.doi.org/10.1007/JHEP06(2012)114}{{\em JHEP} {\bfseries 06}
  (2012) 114},
\href{http://arxiv.org/abs/1204.1698}{{\ttfamily arXiv:1204.1698 [hep-th]}}.

\bibitem{Headrick:2014cta}
M.~Headrick, V.~E. Hubeny, A.~Lawrence, and M.~Rangamani, ``{Causality \&
  holographic entanglement entropy},''
  \href{http://dx.doi.org/10.1007/JHEP12(2014)162}{{\em JHEP} {\bfseries 12}
  (2014) 162},
\href{http://arxiv.org/abs/1408.6300}{{\ttfamily arXiv:1408.6300 [hep-th]}}.

\bibitem{Kachru:2008yh}
S.~Kachru, X.~Liu, and M.~Mulligan, ``{Gravity duals of Lifshitz-like fixed
  points},'' \href{http://dx.doi.org/10.1103/PhysRevD.78.106005}{{\em Phys.
  Rev.} {\bfseries D78} (2008) 106005},
\href{http://arxiv.org/abs/0808.1725}{{\ttfamily arXiv:0808.1725 [hep-th]}}.

\bibitem{Taylor:2008tg}
M.~Taylor, ``{Non-relativistic holography},''
\href{http://arxiv.org/abs/0812.0530}{{\ttfamily arXiv:0812.0530 [hep-th]}}.

\bibitem{Taylor:2015glc}
M.~Taylor, ``{Lifshitz holography},'' {\em Class. Quant. Grav.} {\bfseries 33}
  no.~3, (2016) 033001,
\href{http://arxiv.org/abs/1512.03554}{{\ttfamily arXiv:1512.03554 [hep-th]}}.

\bibitem{Azeyanagi:2009pr}
T.~Azeyanagi, W.~Li, and T.~Takayanagi, ``{On String Theory Duals of
  Lifshitz-like Fixed Points},''
  \href{http://dx.doi.org/10.1088/1126-6708/2009/06/084}{{\em JHEP} {\bfseries
  06} (2009) 084},
\href{http://arxiv.org/abs/0905.0688}{{\ttfamily arXiv:0905.0688 [hep-th]}}.

\bibitem{Solodukhin:2009sk}
S.~N. Solodukhin, ``{Entanglement Entropy in Non-Relativistic Field
  Theories},'' \href{http://dx.doi.org/10.1007/JHEP04(2010)101}{{\em JHEP}
  {\bfseries 04} (2010) 101},
\href{http://arxiv.org/abs/0909.0277}{{\ttfamily arXiv:0909.0277 [hep-th]}}.

\bibitem{Keranen:2011xs}
V.~Ker\"anen, E.~Keski-Vakkuri, and L.~Thorlacius, ``{Thermalization and
  entanglement following a non-relativistic holographic quench},''
  \href{http://dx.doi.org/10.1103/PhysRevD.85.026005}{{\em Phys. Rev.}
  {\bfseries D85} (2012) 026005},
\href{http://arxiv.org/abs/1110.5035}{{\ttfamily arXiv:1110.5035 [hep-th]}}.

\bibitem{Kim:2012nb}
B.~S. Kim, ``{Schr\"odinger Holography with and without Hyperscaling
  Violation},'' \href{http://dx.doi.org/10.1007/JHEP06(2012)116}{{\em JHEP}
  {\bfseries 06} (2012) 116},
\href{http://arxiv.org/abs/1202.6062}{{\ttfamily arXiv:1202.6062 [hep-th]}}.

\bibitem{Alishahiha:2014cwa}
M.~Alishahiha, A.~F. Astaneh, and M.~R.~M. Mozaffar, ``{Thermalization in
  backgrounds with hyperscaling violating factor},''
  \href{http://dx.doi.org/10.1103/PhysRevD.90.046004}{{\em Phys. Rev.}
  {\bfseries D90} no.~4, (2014) 046004},
\href{http://arxiv.org/abs/1401.2807}{{\ttfamily arXiv:1401.2807 [hep-th]}}.

\bibitem{Fonda:2014ula}
P.~Fonda, L.~Franti, V.~Ker\"anen, E.~Keski-Vakkuri, L.~Thorlacius, and
  E.~Tonni, ``{Holographic thermalization with Lifshitz scaling and
  hyperscaling violation},''
  \href{http://dx.doi.org/10.1007/JHEP08(2014)051}{{\em JHEP} {\bfseries 08}
  (2014) 051},
\href{http://arxiv.org/abs/1401.6088}{{\ttfamily arXiv:1401.6088 [hep-th]}}.

\bibitem{Fischetti:2014zja}
S.~Fischetti and D.~Marolf, ``{Complex Entangling Surfaces for AdS and Lifshitz
  Black Holes?},'' \href{http://dx.doi.org/10.1088/0264-9381/31/21/214005}{{\em
  Class. Quant. Grav.} {\bfseries 31} no.~21, (2014) 214005},
\href{http://arxiv.org/abs/1407.2900}{{\ttfamily arXiv:1407.2900 [hep-th]}}.

\bibitem{Hosseini:2015gua}
S.~M. Hosseini and A.~Veliz-Osorio, ``{Entanglement and mutual information in
  2d nonrelativistic field theories},''
\href{http://arxiv.org/abs/1510.03876}{{\ttfamily arXiv:1510.03876 [hep-th]}}.

\bibitem{Singh:2013iba}
H.~Singh, ``{Lifshitz to AdS flow with interpolating $p$-brane solutions},''
  \href{http://dx.doi.org/10.1007/JHEP08(2013)097}{{\em JHEP} {\bfseries 08}
  (2013) 097},
\href{http://arxiv.org/abs/1305.3784}{{\ttfamily arXiv:1305.3784 [hep-th]}}.

\bibitem{Singh:2013pfa}
H.~Singh, ``{Schr\"odinger Spacetimes with Screen and Reduced Entanglement},''
\href{http://arxiv.org/abs/1309.7908}{{\ttfamily arXiv:1309.7908 [hep-th]}}.

\bibitem{Fradkin:2006mb}
E.~Fradkin and J.~E. Moore, ``{Entanglement entropy of 2D conformal quantum
  critical points: hearing the shape of a quantum drum},''
  \href{http://dx.doi.org/10.1103/PhysRevLett.97.050404}{{\em Phys. Rev. Lett.}
  {\bfseries 97} (2006) 050404},
\href{http://arxiv.org/abs/cond-mat/0605683}{{\ttfamily arXiv:cond-mat/0605683
  [cond-mat.str-el]}}.

\bibitem{Hsu:2010ag}
B.~Hsu and E.~Fradkin, ``{Universal Behavior of Entanglement in 2D Quantum
  Critical Dimer Models},''
  \href{http://dx.doi.org/10.1088/1742-5468/2010/09/P09004}{{\em J. Stat.
  Mech.} {\bfseries 1009} (2010) P09004},
\href{http://arxiv.org/abs/1006.1361}{{\ttfamily arXiv:1006.1361
  [cond-mat.stat-mech]}}.

\bibitem{Oshikawa:2010kv}
M.~Oshikawa, ``{Boundary Conformal Field Theory and Entanglement Entropy in
  Two-Dimensional Quantum Lifshitz Critical Point},''
\href{http://arxiv.org/abs/1007.3739}{{\ttfamily arXiv:1007.3739
  [cond-mat.stat-mech]}}.

\bibitem{InglisMelko}
S.~Inglis and R.~G. Melko, ``{Entanglement at a two-dimensional quantum
  critical point: a T = 0 projector quantum Monte Carlo study},'' {\em New
  Journal of Physics} {\bfseries 15} no.~7, (2013) 073048.
  \url{http://stacks.iop.org/1367-2630/15/i=7/a=073048}.

\bibitem{Horava:2009vy}
P.~Ho\v{r}ava and C.~M. Melby-Thompson, ``{Anisotropic Conformal Infinity},''
  \href{http://dx.doi.org/10.1007/s10714-010-1117-y}{{\em Gen. Rel. Grav.}
  {\bfseries 43} (2011) 1391--1400},
\href{http://arxiv.org/abs/0909.3841}{{\ttfamily arXiv:0909.3841 [hep-th]}}.

\bibitem{Ross:2009ar}
S.~F. Ross and O.~Saremi, ``{Holographic stress tensor for non-relativistic
  theories},'' \href{http://dx.doi.org/10.1088/1126-6708/2009/09/009}{{\em
  JHEP} {\bfseries 0909} (2009) 009},
\href{http://arxiv.org/abs/0907.1846}{{\ttfamily arXiv:0907.1846 [hep-th]}}.

\bibitem{Ross:2011gu}
S.~F. Ross, ``{Holography for asymptotically locally Lifshitz spacetimes},''
  \href{http://dx.doi.org/10.1088/0264-9381/28/21/215019}{{\em
  Class.Quant.Grav.} {\bfseries 28} (2011) 215019},
\href{http://arxiv.org/abs/1107.4451}{{\ttfamily arXiv:1107.4451 [hep-th]}}.

\bibitem{Chemissany:2014xsa}
W.~Chemissany and I.~Papadimitriou, ``{Lifshitz holography: The whole
  shebang},'' \href{http://dx.doi.org/10.1007/JHEP01(2015)052}{{\em JHEP}
  {\bfseries 01} (2015) 052},
\href{http://arxiv.org/abs/1408.0795}{{\ttfamily arXiv:1408.0795 [hep-th]}}.

\bibitem{Andrade:2014iia}
T.~Andrade, C.~Keeler, A.~Peach, and S.~F. Ross, ``{Schr\"odinger holography
  for z $\leq$ 2},''
  \href{http://dx.doi.org/10.1088/0264-9381/32/3/035015}{{\em Class. Quant.
  Grav.} {\bfseries 32} no.~3, (2015) 035015},
\href{http://arxiv.org/abs/1408.7103}{{\ttfamily arXiv:1408.7103 [hep-th]}}.

\bibitem{Andrade:2014kba}
T.~Andrade, C.~Keeler, A.~Peach, and S.~F. Ross, ``{Schr\"odinger holography
  with z = 2},'' \href{http://dx.doi.org/10.1088/0264-9381/32/8/085006}{{\em
  Class. Quant. Grav.} {\bfseries 32} no.~8, (2015) 085006},
\href{http://arxiv.org/abs/1412.0031}{{\ttfamily arXiv:1412.0031 [hep-th]}}.

\bibitem{Hartong:2014oma}
J.~Hartong, E.~Kiritsis, and N.~A. Obers, ``{Lifshitz space-times for
  Schr\"odinger holography},''
  \href{http://dx.doi.org/10.1016/j.physletb.2015.05.010}{{\em Phys. Lett.}
  {\bfseries B746} (2015) 318--324},
\href{http://arxiv.org/abs/1409.1519}{{\ttfamily arXiv:1409.1519 [hep-th]}}.

\bibitem{Hartong:2014pma}
J.~Hartong, E.~Kiritsis, and N.~A. Obers, ``{Schr\"odinger Invariance from
  Lifshitz Isometries in Holography and Field Theory},''
  \href{http://dx.doi.org/10.1103/PhysRevD.92.066003}{{\em Phys. Rev.}
  {\bfseries D92} (2015) 066003},
\href{http://arxiv.org/abs/1409.1522}{{\ttfamily arXiv:1409.1522 [hep-th]}}.

\bibitem{Christensen:2013lma}
M.~H. Christensen, J.~Hartong, N.~A. Obers, and B.~Rollier, ``{Torsional
  Newton-Cartan Geometry and Lifshitz Holography},''
  \href{http://dx.doi.org/10.1103/PhysRevD.89.061901}{{\em Phys. Rev.}
  {\bfseries D89} (2014) 061901},
\href{http://arxiv.org/abs/1311.4794}{{\ttfamily arXiv:1311.4794 [hep-th]}}.

\bibitem{Christensen:2013rfa}
M.~H. Christensen, J.~Hartong, N.~A. Obers, and B.~Rollier, ``{Boundary
  Stress-Energy Tensor and Newton-Cartan Geometry in Lifshitz Holography},''
  \href{http://dx.doi.org/10.1007/JHEP01(2014)057}{{\em JHEP} {\bfseries 01}
  (2014) 057},
\href{http://arxiv.org/abs/1311.6471}{{\ttfamily arXiv:1311.6471 [hep-th]}}.

\bibitem{Hoyos:2010at}
C.~Hoyos and P.~Koroteev, ``{On the Null Energy Condition and Causality in
  Lifshitz Holography},'' \href{http://dx.doi.org/10.1103/PhysRevD.82.109905,
  10.1103/PhysRevD.82.084002}{{\em Phys. Rev.} {\bfseries D82} (2010) 084002},
  \href{http://arxiv.org/abs/1007.1428}{{\ttfamily arXiv:1007.1428 [hep-th]}}.
[Erratum: Phys. Rev.D82,109905(2010)].

\bibitem{Engelhardt:2015dta}
N.~Engelhardt and S.~Fischetti, ``{Covariant Constraints on Hole-ography},''
  \href{http://dx.doi.org/10.1088/0264-9381/32/19/195021}{{\em Class. Quant.
  Grav.} {\bfseries 32} no.~19, (2015) 195021},
\href{http://arxiv.org/abs/1507.00354}{{\ttfamily arXiv:1507.00354 [hep-th]}}.

\bibitem{Wall:2012uf}
A.~C. Wall, ``{Maximin Surfaces, and the Strong Subadditivity of the Covariant
  Holographic Entanglement Entropy},''
  \href{http://dx.doi.org/10.1088/0264-9381/31/22/225007}{{\em Class. Quant.
  Grav.} {\bfseries 31} no.~22, (2014) 225007},
\href{http://arxiv.org/abs/1211.3494}{{\ttfamily arXiv:1211.3494 [hep-th]}}.

\bibitem{Hubeny:2013gba}
V.~E. Hubeny, M.~Rangamani, and E.~Tonni, ``{Global properties of causal wedges
  in asymptotically AdS spacetimes},''
  \href{http://dx.doi.org/10.1007/JHEP10(2013)059}{{\em JHEP} {\bfseries 1310}
  (2013) 059},
\href{http://arxiv.org/abs/1306.4324}{{\ttfamily arXiv:1306.4324 [hep-th]}}.

\bibitem{Wald:1984rg}
R.~M. Wald, {\em {General Relativity}}.
\newblock
1984.
\newblock

\bibitem{Kelly:2013aja}
W.~R. Kelly and A.~C. Wall, ``{Coarse-grained entropy and causal holographic
  information in AdS/CFT},''
  \href{http://dx.doi.org/10.1007/JHEP03(2014)118}{{\em JHEP} {\bfseries 03}
  (2014) 118},
\href{http://arxiv.org/abs/1309.3610}{{\ttfamily arXiv:1309.3610 [hep-th]}}.

\bibitem{Gentle:2013fma}
S.~A. Gentle and M.~Rangamani, ``{Holographic entanglement and causal
  information in coherent states},''
  \href{http://dx.doi.org/10.1007/JHEP01(2014)120}{{\em JHEP} {\bfseries 01}
  (2014) 120},
\href{http://arxiv.org/abs/1311.0015}{{\ttfamily arXiv:1311.0015 [hep-th]}}.

\bibitem{Bousso:2002ju}
R.~Bousso, ``{The Holographic principle},''
  \href{http://dx.doi.org/10.1103/RevModPhys.74.825}{{\em Rev. Mod. Phys.}
  {\bfseries 74} (2002) 825--874},
\href{http://arxiv.org/abs/hep-th/0203101}{{\ttfamily arXiv:hep-th/0203101
  [hep-th]}}.

\bibitem{Lewkowycz:2013nqa}
A.~Lewkowycz and J.~Maldacena, ``{Generalized gravitational entropy},''
  \href{http://dx.doi.org/10.1007/JHEP08(2013)090}{{\em JHEP} {\bfseries 08}
  (2013) 090},
\href{http://arxiv.org/abs/1304.4926}{{\ttfamily arXiv:1304.4926 [hep-th]}}.

\bibitem{Bagchi:2014iea}
A.~Bagchi, R.~Basu, D.~Grumiller, and M.~Riegler, ``{Entanglement entropy in
  Galilean conformal field theories and flat holography},''
  \href{http://dx.doi.org/10.1103/PhysRevLett.114.111602}{{\em Phys. Rev.
  Lett.} {\bfseries 114} no.~11, (2015) 111602},
\href{http://arxiv.org/abs/1410.4089}{{\ttfamily arXiv:1410.4089 [hep-th]}}.

\bibitem{Hosseini:2015uba}
S.~M. Hosseini and A.~Veliz-Osorio, ``{Gravitational anomalies, entanglement
  entropy, and flat-space holography},''
\href{http://arxiv.org/abs/1507.06625}{{\ttfamily arXiv:1507.06625 [hep-th]}}.

\bibitem{Castro:2015csg}
A.~Castro, D.~M. Hofman, and N.~Iqbal, ``{Entanglement Entropy in Warped
  Conformal Field Theories},''
\href{http://arxiv.org/abs/1511.00707}{{\ttfamily arXiv:1511.00707 [hep-th]}}.

\bibitem{Haehl:2015rza}
F.~M. Haehl, ``{Comments on universal properties of entanglement entropy and
  bulk reconstruction},'' \href{http://dx.doi.org/10.1007/JHEP10(2015)159}{{\em
  JHEP} {\bfseries 10} (2015) 159},
\href{http://arxiv.org/abs/1508.00766}{{\ttfamily arXiv:1508.00766 [hep-th]}}.

\bibitem{deBoer:2015kda}
J.~de~Boer, M.~P. Heller, R.~C. Myers, and Y.~Neiman, ``{Entanglement
  Holography},''
\href{http://arxiv.org/abs/1509.00113}{{\ttfamily arXiv:1509.00113 [hep-th]}}.

\bibitem{Gao:2000ga}
S.~Gao and R.~M. Wald, ``{Theorems on gravitational time delay and related
  issues},'' \href{http://dx.doi.org/10.1088/0264-9381/17/24/305}{{\em Class.
  Quant. Grav.} {\bfseries 17} (2000) 4999--5008},
\href{http://arxiv.org/abs/gr-qc/0007021}{{\ttfamily arXiv:gr-qc/0007021
  [gr-qc]}}.

\bibitem{Lieb:1972wy}
E.~H. Lieb and D.~W. Robinson, ``{The finite group velocity of quantum spin
  systems},''
\href{http://dx.doi.org/10.1007/BF01645779}{{\em Commun. Math. Phys.}
  {\bfseries 28} (1972) 251--257}.

\bibitem{Hofman:2014loa}
D.~M. Hofman and B.~Rollier, ``{Warped Conformal Field Theory as Lower Spin
  Gravity},'' \href{http://dx.doi.org/10.1016/j.nuclphysb.2015.05.011}{{\em
  Nucl. Phys.} {\bfseries B897} (2015) 1--38},
\href{http://arxiv.org/abs/1411.0672}{{\ttfamily arXiv:1411.0672 [hep-th]}}.

\end{thebibliography}

\providecommand{\href}[2]{#2}\begingroup\raggedright\endgroup

\end{document}